\documentclass[useAMS,usenatbib]{mn2e}
\usepackage{footnote}
\usepackage{graphicx}
\usepackage{amsmath}
\usepackage{natbib}
\usepackage{array}
\usepackage{color}
\usepackage{url}
\definecolor{titlecol}{rgb}{0,0,0}
\def\changed    {\color{titlecol} }
\definecolor{change}{rgb}{0,0,0}
\def\newchange {\color{change} }
\def\refchange {\color{ref} }
\definecolor{ref}{rgb}{0,0,0}
\def\tworef {\color{tref} }
\definecolor{tref}{rgb}{0,0,0}

\def\starpy {\textsc{starpy}}

\begin{document}
\title[The Star Formation History of the Green Valley]{Galaxy Zoo: Evidence for Diverse Star Formation Histories through the Green Valley}
\author[Smethurst et al. 2014]{R. ~J. ~Smethurst,$^{1}$ C. ~J. ~Lintott,$^{1}$ B. ~D. ~Simmons,$^{1}$ K. ~Schawinski,$^{2}$ \newauthor  P. ~J. ~Marshall,$^{3,1}$ S. ~Bamford,$^{4}$ L. Fortson,$^{5}$ S. ~Kaviraj,$^{6}$ K.~L.~Masters,$^{7}$ \newauthor  T. ~Melvin,$^{7}$  R. C. ~Nichol,$^{7}$  R. ~A. ~Skibba,$^{8}$ K. ~W. ~Willett$^{5}$ 
\\ $^1$ Oxford Astrophysics, Department of Physics, University of Oxford, Denys Wilkinson Building, Keble Road, Oxford, OX1 3RH, UK 
\\ $^2$ Institute for Astronomy, Department of Physics, ETH Zurich, Wolfgang-Pauli Strasse 27, CH-8093 Zurich, Switzerland 
\\ $^3$ Kavli Institute for Particle Astrophysics and Cosmology, Stanford University, 452 Lomita Mall, Stanford, CA 95616, USA
\\ $^4$ School of Physics and Astronomy, The University of Nottingham, University Park, Nottingham, NG7 2RD, UK
\\ $^5$ School of Physics and Astronomy, University of Minnesota, 116 Church St SE, Minneapolis, MN 55455, USA
\\ $^6$ Centre for Astrophysics Research, University of Hertfordshire, College Lane, Hatfield, Hertfordshire, AL10 9AB, UK
\\ $^7$ Institute of Cosmology and Gravitation, University of Portsmouth, Dennis Sciama Building, Barnaby Road, Portsmouth, PO1 3FX, UK 
\\ $^8$ Center for Astrophysics and Space Sciences, University of California San Diego, 9500 Gilman Drive, La Jolla, CA 92093, USA
\\
\\Accepted 2015 January 22.  Received 2015 January 14; in original form 2014 September 17
}

\maketitle

\begin{abstract}
{\refchange Does galaxy evolution proceed through the green valley via multiple pathways or as a single population? Motivated by recent results  highlighting radically different evolutionary pathways between early- and late-type galaxies, we present results from a simple Bayesian approach to this problem wherein we model the star formation history (SFH) of a galaxy with two parameters, $[t, \tau]$ and compare the predicted and observed optical and near-ultraviolet colours. We use a novel method to investigate the morphological differences between the most probable SFHs for both disc-like and smooth-like populations of galaxies, by using a sample of $126,316$ galaxies $(0.01 < z < 0.25)$ with probabilistic estimates of morphology from Galaxy Zoo. We find a clear difference between the quenching timescales preferred by smooth- and disc-like galaxies, with three possible routes through the green valley dominated by smooth- (rapid timescales, attributed to major mergers), intermediate- (intermediate timescales, attributed to minor mergers and galaxy interactions) and disc-like (slow timescales, attributed to secular evolution) galaxies. We hypothesise that morphological changes occur in systems which have undergone quenching with an exponential timescale $\tau < 1.5~\rm{Gyr}$, in order for the evolution of galaxies in the green valley to match the ratio of smooth to disc galaxies observed in the red sequence. These rapid timescales are instrumental in the formation of the red sequence at earlier times; however we find that galaxies currently passing through the green valley typically do so at intermediate timescales.\footnotemark[1]
}
\end{abstract}

\footnotetext[1]{This investigation has been made possible by the participation of more than 250,000 users in the Galaxy Zoo project. Their contributions are individually acknowledged at \url{http://authors.galaxyzoo.org}}

\section{Introduction}

Previous large scale surveys of galaxies have revealed a bimodality in the colour-magnitude diagram (CMD) with two distinct populations; one at relatively low mass, with blue optical colours and another at relatively high mass, with red optical colours \citep{Baldry04, Baldry06, Willmer06, BLB08, Brammer09}. These populations were dubbed the `blue cloud' and `red sequence' respectively {\refchange \citep{Chester64, BLE92, Driver06, Faber07}}. The Galaxy Zoo project \citep{Lintott11}, which produced morphological classifications for a million galaxies, helped to confirm that this bimodality is not entirely morphology driven {\refchange \citep{Strat01, Salim07, Sch07, CHV08, Bamford09, Skibba09}, detecting larger fractions of spiral galaxies in the red sequence \citep{Masters10} and elliptical galaxies in the blue cloud \citep{Sch09} than had previously been detected. }

\begin{figure*}
\centering{
\includegraphics[width=\textwidth]{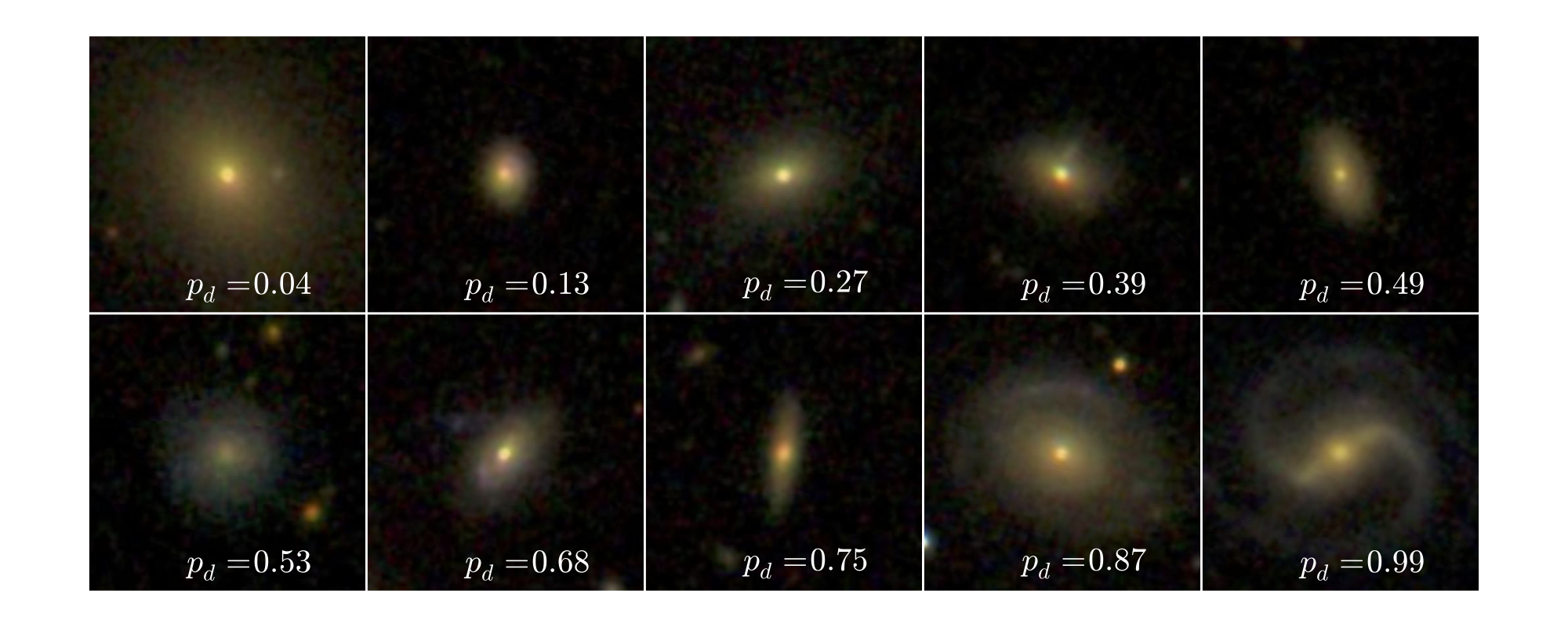}}
\caption{{\newchange Randomly selected SDSS \emph{gri} composite images showing the continuous probabilistic nature of the Galaxy Zoo sample from a redshift range $0.070 < z < 0.075$. The debiased disc vote fraction (see \citealt{GZ2}) for each galaxy is shown. The scale for each image is $0.099~\rm{arcsec/pixel}$.}}
\label{mosaic}
\end{figure*}

The sparsely populated colour space between these two populations, the so-called `green valley', provides clues to the nature and duration of galaxies' transitions from blue to red. This transition must occur on rapid timescales, otherwise there would be an accumulation of galaxies residing in the green valley, rather than an accumulation in the red sequence as is observed \citep{Arnouts07, Martin07}. Green valley galaxies have therefore long been thought of as the `crossroads' of galaxy evolution, a transition population between the two main galactic stages of the star forming blue cloud and the `dead' red sequence \citep{Bell04, Wyder07, Schim07, Martin07, Faber07, Mendez11, Gonc12, Sch2014, Pan14}.

The intermediate colours of these green valley galaxies have been interpreted as evidence for recent quenching (suppression) of star formation \citep{Salim07}. Star forming galaxies are observed to lie on a well defined mass-SFR relation, however quenching a galaxy causes it to depart from this relation (\citealt{Noeske07, Peng}; see Figure~\ref{sfr_mass_sub})

By studying the galaxies which  have just left this mass-SFR relation, we can probe the quenching mechanisms by which this occurs. There have been many previous theories for the initial triggers of these quenching mechanisms, including negative feedback from AGN {\refchange \citep{diMatteo05, Martin07, Nandra07, Sch07}, mergers \citep{Darg10a, Cheung12, Barro13}, supernovae winds \citep{MFB12}, cluster interactions \citep{Coil08, Mendez11, Fang13} and secular evolution \citep{Masters10, Masters11, Mendez11}.} By investigating the \emph{amount} of quenching that has occurred in the blue cloud, green valley and red sequence; and by comparing the amount across these three populations, we can apply some constraints to these theories. 

We have been motivated by a recent result suggesting two contrasting evolutionary pathways through the green valley by different morphological types (\citealt{Sch2014}, hereafter S14), specifically that late-type galaxies quench very slowly and form a nearly static disc population in the green valley, whereas early-type galaxies quench very rapidly, transitioning through the green valley and onto the red sequence in $\sim 1$~Gyr \citep{Wong12}. That study used a toy model to examine quenching across the green valley. Here we implement a novel method utilising Bayesian statistics (for a comprehensive overview of Bayesian statistics see either \citealt{MacKay} or \citealt{Sivia}) in order to find the most likely model description of the star formation histories of galaxies in the three populations. This method also enables a direct comparison with our current understanding of galaxy evolution from stellar population synthesis (SPS, see section~\ref{models}) models.

\begin{table*}
\caption{Table showing the decomposition of the GZ2 sample by galaxy type into the subsets of the colour-magnitude diagram.}
\begin{tabular*}{0.9\textwidth}{r @{\extracolsep{\fill}}cccc}
\hline
\begin{tabular}[c]{@{}c@{}} {\color{white} -} \\ {\color{white} -}  \end{tabular} & All                                                      & Red Sequence                                              & Green Valley                                              & Blue Cloud \\  \hline 
Smooth-like ($p_s > 0.5$)        & \begin{tabular}[c]{@{}c@{}}42453\\ (33.6\%)\end{tabular} & \begin{tabular}[c]{@{}c@{}}17424\\ (61.9\%)\end{tabular}  & \begin{tabular}[c]{@{}c@{}}10687\\ (44.6\%)\end{tabular}   & \begin{tabular}[c]{@{}c@{}}14342\\ (19.3\%)\end{tabular}  \\ 
Disc-like ($p_d > 0.5$)          & \begin{tabular}[c]{@{}c@{}}83863\\ (80.7\%)\end{tabular} & \begin{tabular}[c]{@{}c@{}}10722\\ (38.1\%)\end{tabular}   & \begin{tabular}[c]{@{}c@{}}13257\\ (55.4\%)\end{tabular}  & \begin{tabular}[c]{@{}c@{}}59884\\ (47.4\%)\end{tabular}  \\
Early-type ($p_s \geq 0.8$) & \begin{tabular}[c]{@{}c@{}}10517\\ (8.3\%)\end{tabular}  & \begin{tabular}[c]{@{}c@{}}5337\\ (18.9\%)\end{tabular}    & \begin{tabular}[c]{@{}c@{}}2496\\ (10.4\%)\end{tabular}    & \begin{tabular}[c]{@{}c@{}}2684\\ (3.6\%)\end{tabular}    \\
Late-type ($p_s \geq 0.8$)  & \begin{tabular}[c]{@{}c@{}}51470\\ (40.9\%)\end{tabular} & \begin{tabular}[c]{@{}c@{}}4493\\ (15.9\%)\end{tabular}    & \begin{tabular}[c]{@{}c@{}}6817\\ (28.5\%)\end{tabular}    & \begin{tabular}[c]{@{}c@{}}40430\\ (54.4\%)\end{tabular}  \\ \hline
\textbf{Total}                       & \begin{tabular}[c]{@{}c@{}}\textbf{126316} \\ (100.0\%)\end{tabular}                                                & \begin{tabular}[c]{@{}c@{}}28146 \\ (22.3\%)\end{tabular} & \begin{tabular}[c]{@{}c@{}}23944 \\ (18.9\%)\end{tabular} & \begin{tabular}[c]{@{}c@{}}74226 \\ (58.7\%)\end{tabular} \\\hline
\end{tabular*}
\label{subs}
\end{table*}

Through this approach, we aim to determine the following:
\begin{enumerate}
\item What previous star formation history (SFH) causes a galaxy to reside in the green valley at the current epoch?
\item Is the green valley a transitional or static population? 
\item If the green valley is a transitional population, how many routes through it are there? 
\item Are there morphology-dependent differences between these routes through the green valley? 
\end{enumerate}

This paper proceeds as follows. Section~\ref{data} contains a description of the sample data, which is used in the Bayesian analysis of an exponentially declining star formation history model, all described in Section~\ref{models}. Section~\ref{results} contains the results produced by this analysis, with Section~\ref{diss} providing a detailed discussion of the results obtained. We also summarise our findings in Section~\ref{conc}. The zero points of all \emph{ugriz} magnitudes are in the AB system and where necessary we adopt the WMAP Seven-Year Cosmological parameters \citep{WMAP} with $(\Omega_m, \Omega_{\lambda}, h) = (0.26, 0.73, 0.71)$. 

\section{Data}\label{data}
\subsection{Multi-wavelength data}\label{multi}
The galaxy sample is compiled from publicly available optical data from the Sloan Digitial Sky Survey (SDSS; \citealt{York00}) Data Release 8 \citep{Aihara11}. Near-ultraviolet (NUV) photometry was obtained from the Galaxy Evolution Explorer (GALEX; \citealt{Martin05}) and was matched with a search radius of $1''$ in right ascension and declination. 

{\refchange Observed optical and ultraviolet fluxes are corrected for galactic extinction \citep{Oh11} by applying the \citet*{Cardelli89} law, giving an typical correction of $u-r \sim 0.05$. We also adopt k-corrections to $z=0.0$ and obtain absolute magnitudes from the NYU-VAGC \citep{Blanton05, Pad08, BR07}, giving a typical $u-r$ correction of $\sim 0.15$ mag. The change in the $u-r$ colour due to both corrections therefore ranges from $\Delta u-r \sim 0.2$ at low redshift, increasing up to $\Delta u-r \sim 1.0$ at $z \sim 0.25$, which is consistent with the expected k-corrections shown in Figure 15 of \citet{BR07}. These corrections were calculated by \citet{Bamford09} for the entire Galaxy Zoo sample. These corrections are a crucial aspect of this work since a $\Delta u-r \sim 1.0$ can cause a galaxy to cross the definition between blue cloud, green valley and red sequence.}

We obtained star formation rates and stellar masses from the MPA-JHU catalog (\citealt{Kauff03, Brinch04}; average values, \textsc{AVG}, corrected for aperture and extinction), which are in turn calculated from the SDSS spectra and photometry. 

We further select a sub-sample with detailed morphological classifications, as described below, {\refchange to give a volume limited sample in the redshift range $0.01 < z < 0.25$}.

\subsection{Galaxy Zoo 2 Morphological classifications}\label{class}

\begin{figure}
\centering{
\includegraphics[width=0.48\textwidth]{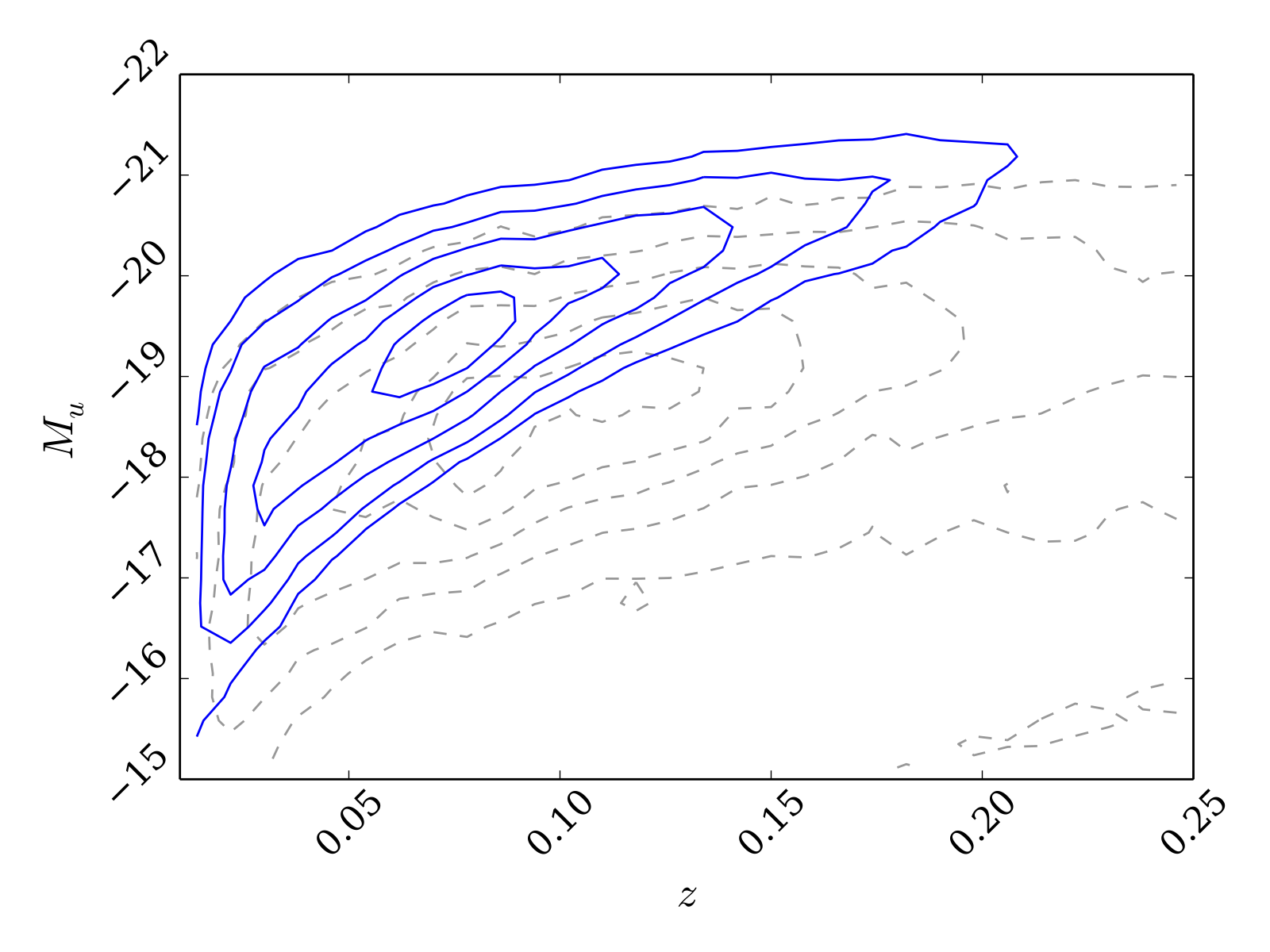}}
\caption{{\refchange Absolute $u$-band magnitude against redshift for the whole of SDSS (grey dashed lines) in comparison to the GZ2 subsample (blue solid lines). Typical Milky Way $L_*$ galaxies with $M_u \sim -20.5$ are still included in the GZ2 subsample out to the highest redshift of $z \sim 0.25$.}}
\label{complete}
\end{figure}

In this investigation we use visual classifications of galaxy morphologies from the Galaxy Zoo 2\footnote{\url{http://zoo2.galaxyzoo.org/}} citizen science project \citep{GZ2}, which obtains multiple independent classifications for each galaxy image; the full question tree for each image is shown in Figure 1 of \citealt{GZ2}.  

The Galaxy Zoo 2 (GZ2) project consists of $304, 022$ images from the SDSS DR8 (a subset of those classified in Galaxy Zoo 1; GZ1) all classified by \emph{at least} 17 independent users, with the mean number of classifications standing at $\sim42$. The GZ2 sample is more robust than the GZ1 sample and provides more detailed morphological classifications, including features such as bars, the number of spiral arms and the ellipticity of smooth galaxies. It is for these reasons we use the GZ2 sample, as opposed to the GZ1, allowing for further investigation of specific galaxy classes in the future (see Section~\ref{future}). The only selection that was made on the sample was to remove objects considered to be stars, artefacts or merging pairs by the users (i.e. with $p_{star/artefact} ~\geq~ 0.8$ or $p_{merger} ~\geq 0.420$; see \citealt{GZ2} Table 3 and discussion for details of this fractional limit). Further to this, we required NUV photometry from the GALEX survey, within which $\sim42\%$ of the GZ2 sample were observed, giving a total sample size of $126, 316$ galaxies. {\refchange The completeness of this subsample of GZ2 matched to GALEX is shown in Figure~\ref{complete} with the $u$-band absolute magnitude against redshift for this sample compared with the SDSS data set. Typical Milky Way $L_*$ galaxies with $M_u \sim -20.5$ are still included in the GZ2 subsample out to the highest redshift of $z \sim 0.25$; however dwarf and lower mass galaxies are only detected at the lowest redshifts.}

\begin{figure}
\centering{
\includegraphics[width=0.48\textwidth]{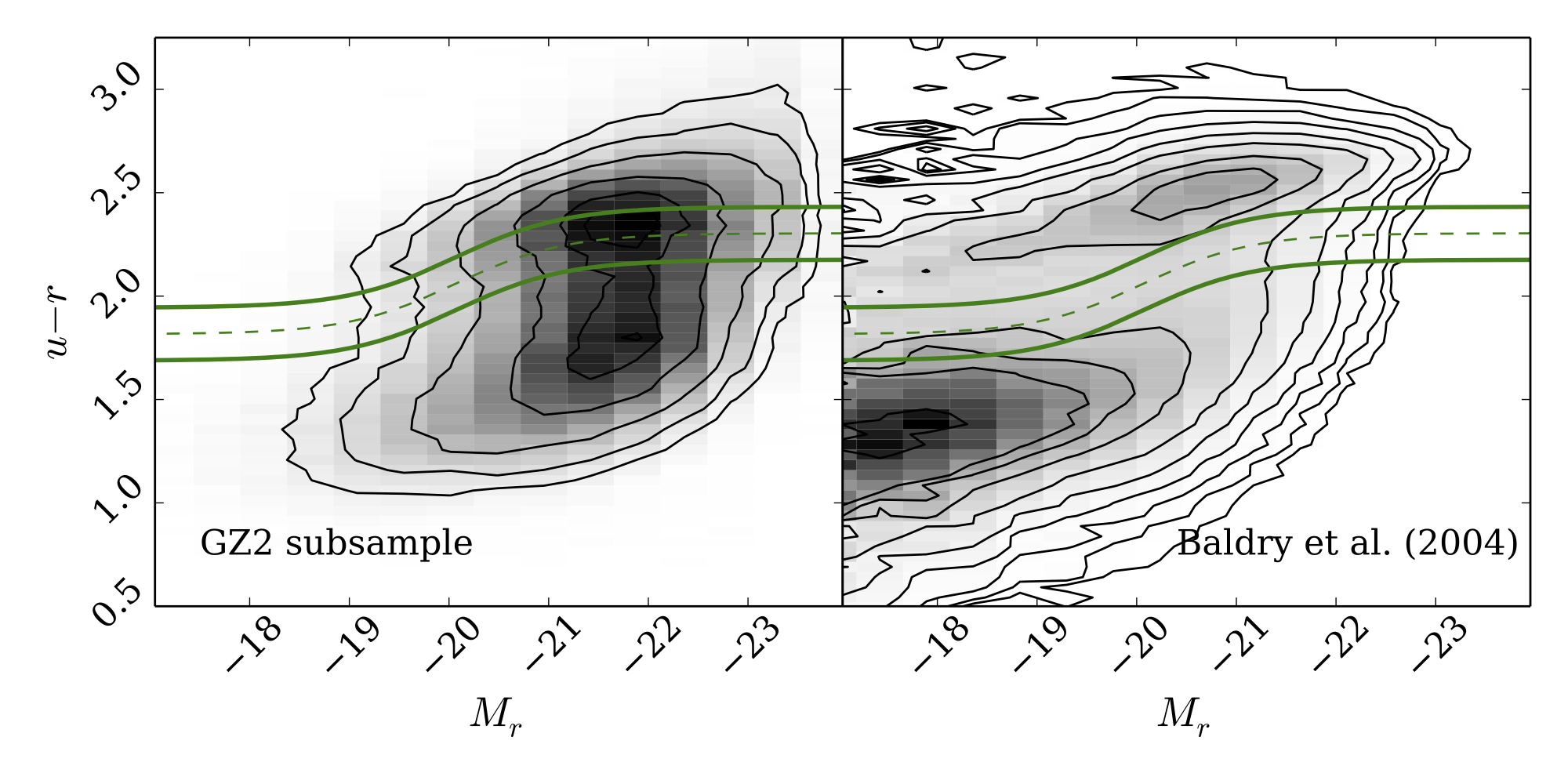}}
\caption{Colour-magnitude diagram for the Galaxy Zoo 2 subsample (left) and the SDSS sample from \citet{Baldry04} with the definition between the blue cloud and the red sequence from \citet{Baldry04} with the dashed line, as defined in Equation~\ref{eqgv}. The solid lines show $\pm 1\sigma$ either side of this definition; any galaxy within the boundary of these two solid lines is considered a green valley galaxy. {\refchange The lack of red sequence galaxies due to the necessity for NUV GALEX colours skews the apparent location of the green valley, therefore a literature definition of the green valley is used to ensure comparisons can be made with other works.}}
\label{CMGV}
\end{figure}

The first task of GZ2 asks users to choose whether a galaxy is mostly smooth, is featured and/or has a disc or is a star/artefact. Unlike other tasks further down in the decision tree, every user who classifies a galaxy image will complete this task (others, such as whether the galaxy has a bar, is dependent on a user having first classified it as a featured galaxy). Therefore we have the most statistically robust classifications at this level.

The classifications from users produces a vote fraction for each galaxy (the debiased fractions calculated by \citet{GZ2} were used in this investigation); for example if 80 of 100 people thought a galaxy was disc shaped, whereas 20 out of 100 people thought the same galaxy was smooth in shape (i.e. elliptical), that galaxy would have vote fractions $p_{s} = 0.2$ and $p_{d} = 0.8$. In this example this galaxy would be included in the \emph{`clean'} disc sample ($p_d \geq 0.8$) according to \cite{GZ2} and would be considered a late-type galaxy. {\changed All previous Galaxy Zoo projects have incorporated extensive analysis of volunteer classifications to measure classification accuracy and bias, and compute user weightings (for a detailed description of debiasing and consistency-based user weightings, see either Section 3 of \citealt{Lintott09} or Section 3 of \citealt{GZ2}). }

{\changed The classifications are highly accurate and provide a continuous scale of morphological features, as shown in Figure~\ref{mosaic}, rather than a simple binary classification separating elliptical and disc galaxies. These classifications allow each galaxy to be considered as a probabilistic object with both bulge and disc components.} For the first time, we incorporate this advantage of the GZ classifications into a large statistical analysis of how elliptical and disc galaxies differ in their SFHs.

\subsection{Defining the Green Valley}\label{defGV}

To define which of the sample of $126, 316$ galaxies were in the green valley, {\changed we looked to previous definitions in the literature defining the separation between the red sequence and blue cloud to ensure comparisons can be made with other works. \citet{Baldry04} used a large sample of local galaxies from the SDSS to trace this bimodality by fitting double Gaussians to the colour magnitude diagram without cuts in morphology.} Their relation is defined in their Equation 11 as:
\begin{equation}\label{eqgv}
C'_{ur}(M_{r}) = 2.06 - 0.244 \tanh \left( \frac{M_r + 20.07}{1.09}\right)
\end{equation}
and is shown in Figure~\ref{CMGV} by the dashed line in comparison to both the GZ2 subsample (left) and the SDSS data from \cite{Baldry04}. {\newchange This ensures that the definition of the green valley used is derived from a complete sample, rather than from our sample that is dominated by blue galaxies due to the necessity for NUV photometry.} Any galaxy within $\pm 1\sigma$ of this relationship, shown by the solid lines in Figure~\ref{CMGV}, is therefore considered a green valley galaxy. The decomposition of the sample into red sequence, green valley and blue cloud galaxies is shown in Table~\ref{subs} along with further subsections by galaxy type. This table also lists the definitions we adopt henceforth for early-type ($p_s~ \geq~0.8$), late-type ($p_d~ \geq~0.8$), smooth-like ($p_s~ >~0.5$) and disc-like ($p_d~ >~0.5$) galaxies.

\section{Models}\label{models}
In the following section, the quenched SFH models are described in Section ~\ref{qmod} and the probabilistic fitting method to the data is described in Section \ref{stats}.
\subsection{Quenching Models}\label{qmod}
 \begin{figure}
\centering{
\includegraphics[width=0.44\textwidth]{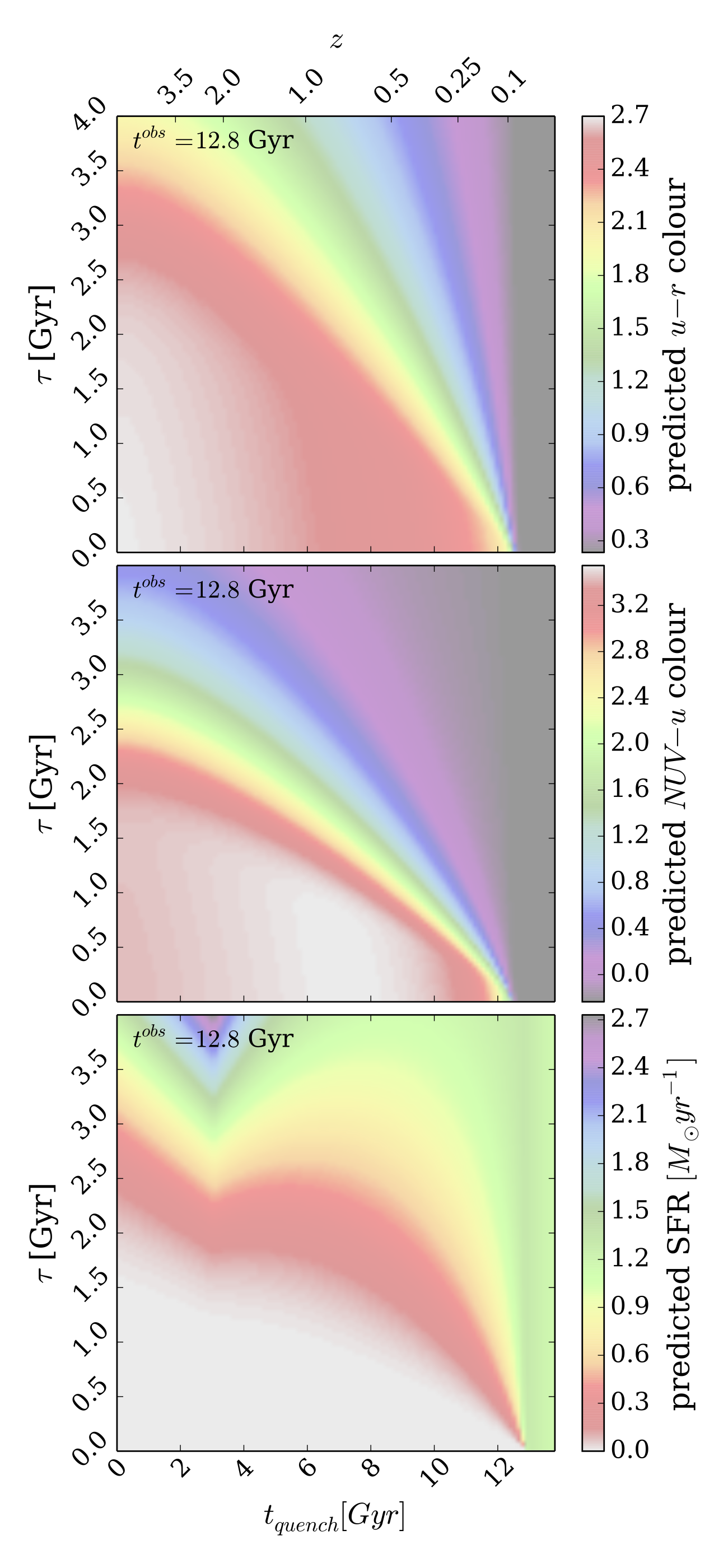}}
\caption{Quenching timescale $\tau$ versus quenching onset time $t$ in all three panels for the quenched SFH models used in ~\starpy. Colour shadings show model predictions of the $u-r$ optical colour (top panel), $NUV-u$ colour (middle panel), and star formation rate in $M_\odot \rm{~yr}^{-1}$ (lower panel), at $t^{obs} = 12.8~\rm{Gyr}$, the mean observed redshift of the GZ2 sample (see Section \ref{qmod}). The combination of optical and NUV colours is a sensitive measure of the $\theta = [t_q, \tau]$ parameter space. Note that all models with $t > 12.8$ \rm{Gyr} are effectively un-quenched. The `kink' in the bottom panel is due to the assumption that the sSFR is constant prior to $t \sim 3~\rm{Gyr}$ ($z\sim 2.2$).}
\label{pred}
\end{figure}

The quenched star formation history (SFH) of a galaxy can be simply modelled as an exponentially declining star formation rate (SFR) across cosmic time ($0 \leq t ~\rm{[Gyr]} \leq 13.8$) as:
\begin{equation}\label{sfh}
SFR =
\begin{cases}
i_{sfr}(t_q) & \text{if } t < t_q \\
i_{sfr}(t_q) \times exp{\left( \frac{-(t-t_{q})}{\tau}\right)} & \text{if } t > t_q 
\end{cases}
\end{equation}
where $t_{q}$ is the onset time of quenching, $\tau$ is the timescale over which the quenching occurs and $i_{sfr}$ is an initial constant star formation rate dependent on $t_q$.  A smaller $\tau$ value corresponds to a rapid quench, whereas a larger $\tau$ value corresponds to a slower quench. 

We assume that all galaxies formed at a time $t=0~\rm{Gyr}$ with an initial burst of star formation. The mass of this initial burst is controlled by the value of the $i_{sfr}$ which is set as the average specific SFR (sSFR) at the time of quenching $t_q$. {\refchange \citet{Peng} defined a relation (their equation 1) between the average sSFR and redshift (cosmic time, $t$) by fitting to measurements of the mean sSFR of blue star forming galaxies from SDSS, zCOSMOS and literature values at increasing redshifts \citep{Elbaz07, Daddi07}:}
\begin{equation}
sSFR(m,t) = 2.5 \left( \frac{m}{10^{10} M_{\odot}} \right)^{-0.1} \left(\frac{t}{3.5  {\refchange ~\rm{Gyr}}}\right)^{-2.2} \rm{Gyr}^{-1}.
\end{equation}
Beyond $z \sim 2$ the characteristic SFR flattens and is roughly constant back to $z\sim6$. The cause for this change is not well understood but can be seen across similar observational data \citep{Peng, Gonzalez, Beth}. Motivated by these observations, the relation defined in \citet{Peng} is taken up to a cosmic time of $t=3~\rm{Gyr}~(z \sim 2.3)$ and prior to this a constant average SFR is assumed (see Figure~\ref{sfr_mass_col}). At the point of quenching, $t_{q}$, the models are defined to have a SFR which lies on this relationship for the sSFR, for a galaxy with mass, $m = 10^{10.27} M_{\odot}$ (the mean mass of the GZ2 sample; see Section~\ref{results} and Figure~\ref{sfr_mass_col}).
  
Under these assumptions the average SFR of our models will result in a lower value than the relation defined in \citet{Peng} at all cosmic times; each galaxy only resides on the `main sequence' at the point of quenching. However galaxies cannot remain on the `main sequence' from early to late times throughout their entire lifetimes given the unphysical stellar masses and SFRs this would result in at the current epoch in the local Universe \citep{Beth, Heinis14}. If we were to include prescriptions for no quenching, starbursts, mergers, AGN etc. into our models we would improve on our reproduction of the average SFR across cosmic time; however we chose to initially focus on the simplest model possible.

Once this evolutionary SFR is obtained, it is convolved with the \citet{BC03} population synthesis models to generate a model SED at each time step. The observed features of galaxy spectra can be modelled using simple stellar population techniques which sum the contributions of individual, coeval, equal-metallicity stars. The accuracy of these predictions depends on the completeness of the input stellar physics. Comprehensive knowledge is therefore required of (i) stellar evolutionary tracks and (ii) the initial mass function (IMF) to synthesise a stellar population accurately. 

These stellar population synthesis (SPS) models are an extremely well explored (and often debated) area of astrophysics \citep{Maraston05, Eminian08, CGW09, Falk09, Chen10, Kriek10, MRC11, Mel12}. In this investigation we chose to utilise the \citet{BC03} \emph{GALEXEV} SPS models, to allow a direct comparison with S14, along with a Chabrier \citep{Chab03} IMF, across a large wavelength range ($0.0091 < ~\lambda~\rm{[\mu m]}~ < 160 $) with solar metallically (m62 in the \citet{BC03} models; hereafter BC03).

Fluxes from stars younger than $3~$Myr in the SPS model are suppressed to mimic the large optical depth of protostars embedded in dusty formation cloud (as in S14), then filter transmission curves are applied to the fluxes to obtain AB magnitudes and therefore colours. {\refchange For a particular galaxy at an observed redshift, $z$, we define the observed time, $t^{obs}$ for that galaxy using the standard cosmological conversion between redshift and time. We utilise the SFH models at this observed time for each individual galaxy to compare the predicted model and observed colours directly.}

Figure~\ref{pred} shows these predicted optical and NUV colours at a time of $t^{obs} = 12.8 ~\rm{Gyr}$ (the average observed time of the Galaxy Zoo 2 sample, $z \sim 0.076$) provided by the exponential SFH model. These predicted colours will be referred to as $d_{c,p}(t_{q}, \tau, t^{obs})$, where c=\{opt,NUV\} and p = predicted. The SFR at a time of $t^{obs}=12.8~\rm{Gyr}$ is also shown in Figure~\ref{pred} to compare how this correlates with the predicted colours. The $u-r$ predicted colour shows an immediate correlation with the SFR, however the $NUV-u$ colour is more sensitive to the value of $\tau$ and so is ideal for tracing any recent star formation in a population . At small $\tau$ (rapid quenching timescales) the $NUV-u$ colour is insensitive to $t_{q}$, whereas at large $\tau$ (slow quenching timescales) the colour is very sensitive to $t_{q}$. Together the two colours are ideal for tracing the effects of $t_{q}$ and $\tau$ in a population. 

{\newchange We stress here that this model is not a fully hydrodynamical simulation, it is a simple model built in order to test the understanding of the evolution of galaxy populations. These models are therefore not expected to accurately determine the SFH of every galaxy in the GZ2 sample, in particular galaxies which have not undergone any quenching. In this case the models described above can only attribute a constant star formation rate to these  unquenched galaxies. In reality, there are many possible forms of SFH that a galaxy can take, a few of which have been investigated in previous literature; starbursts \citep{Canalizo01}, a power law \citep{Glazebrook03}, single stellar populations \citep{Trager00, Sanchez06, Vazdekis10} and metallicity enrichment \citep{deLucia14}. Incorporating these different SFHs along with prescriptions for mergers and a reinvigoration of star formation post quench into our models is a possible future extension to this work once the results of this initial study are well enough understood to permit additional complexity to be added.}

\subsection{Probabilistic Fitting}\label{stats}

In order to achieve robust conclusions we conduct a Bayesian analysis \citep{Sivia, MacKay} of our SFH models in comparison to the observed GZ2 sample data. This approach requires consideration of all possible combinations of $\theta \equiv (t_{q}, \tau)$. Assuming that all galaxies formed at $t=0~\rm{Gyr}$ with an initial burst of star formation, we can assume that the `age' of each galaxy in the GZ2 sample is equivalent to an observed time, $t^{obs}_{k}$ (see Section~\ref{class}). We then use this  `age' to calculate the predicted model colours at this cosmic time for a given combination of $\theta$: $d_{c,p}(\theta_k, t^{obs}_{k})$ for both optical and NUV $(c={opt,NUV})$ colours. We can now directly compare our model colours with the observed GZ2 galaxy colours, so that for a single galaxy $k$ with optical ($u-r$) colour, $d_{opt, k}$ and NUV ($NUV-u$) colour, $d_{NUV,k}$, the {\changed likelihood $P(d_{k}|\theta_k, t^{obs}_{k})$ is}:

\begin{multline}\label{like}
P(d_{k}|\theta_k, t^{obs}_{k}) = \frac{1}{\sqrt{2\pi\sigma_{opt, k}^2}}\frac{1}{\sqrt{2\pi\sigma_{NUV, k}^2}} \\ \exp{\left[ - \frac{(d_{opt, k} - d_{opt, p}(\theta_k, t_{k}^{obs}))^2}{\sigma_{opt, k}^2} \right]} \\ \exp{\left[ - \frac{(d_{NUV, k} - d_{NUV, p}(\theta_k, t_{k}^{obs}))^2}{\sigma_{NUV, k}^2} \right]}.
\end{multline}

We have assumed that $P(d_{opt}|\theta_k, t^{obs}_{k})$ and $P(d_{NUV}|\theta_k, t^{obs}_{k})$ are independent of each other and that the errors on the observed colours are also independent. To obtain the probability of each combination of $\theta$ values \underline{given} the GZ2 data: $P(\theta_k|d_k, t^{obs})$, i.e. how likely is a single SFH model given the observed colours of a single GZ2 galaxy, {\changed we utilise Bayes' theorem}:
{\changed \begin{equation}\label{big}
P(\theta_k|d_k, t^{obs}) = \frac{P(d_k|\theta_k, t^{obs})P(\theta_k)}{\int P(d_k |\theta_k, t^{obs})P(\theta_k) d\theta_k}.
\end{equation}}
{\changed We assume a flat prior on the model parameters so that:
\begin{equation}\label{prior}
P(\theta_k) =
\begin{cases}
1 & \text{if } 0 \leq t_q ~\rm{[Gyr]}~ \leq 13.8 ~  \text{ and } ~ 0 \leq \tau  ~\rm{[Gyr]}~ \leq 4\\
0 & \text{otherwise.} \\
\end{cases}
\end{equation}}

\begin{figure}
\centering{
\includegraphics[width=0.5\textwidth]{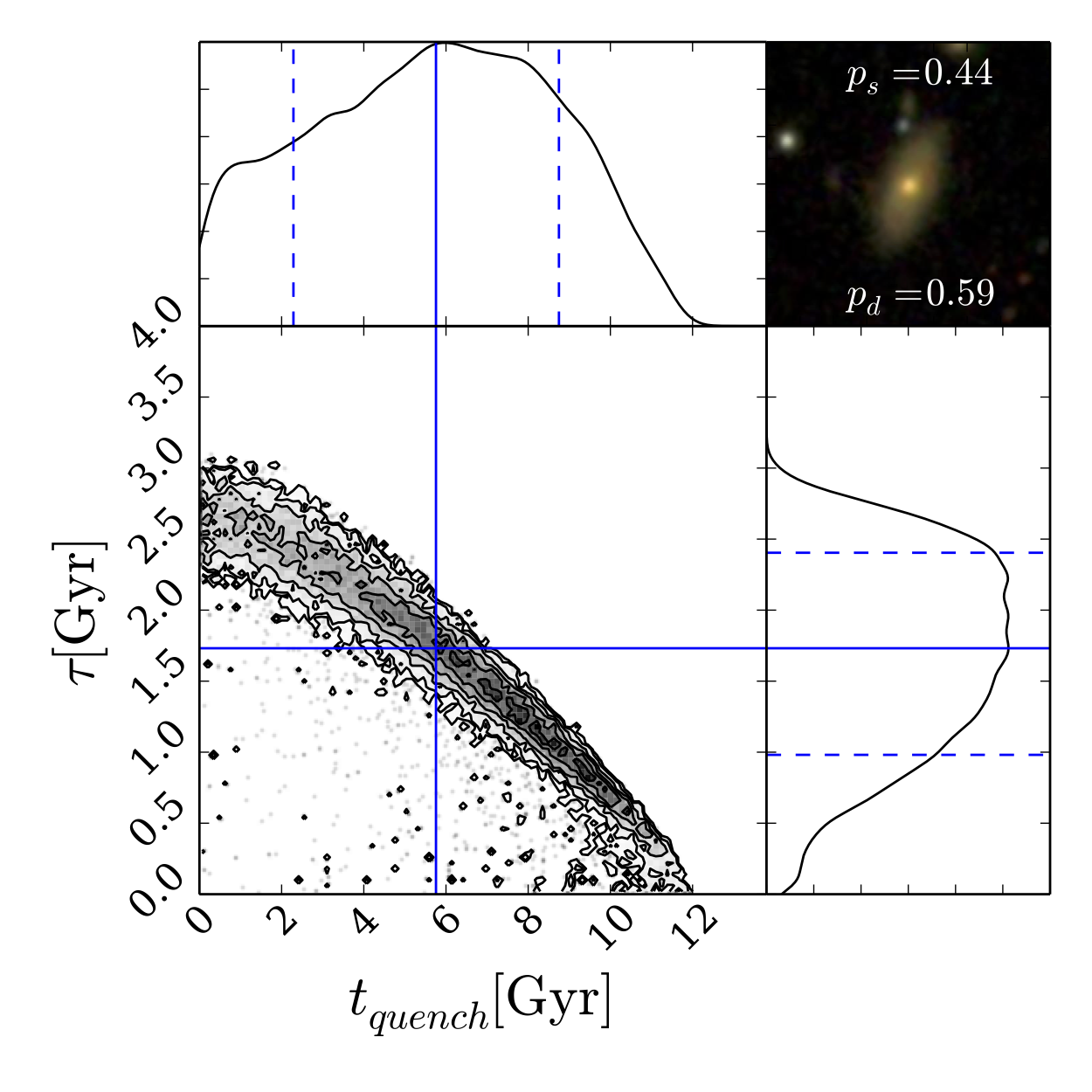}}
\caption{{\changed Example output from \starpy ~for a galaxy within the red sequence. The contours show the positions of the `walkers' in the Markov Chain (which are analogous to the areas of high probability) for the quenching models described by $\theta = [t_q, \tau]$ and the histograms show the 1D projection along each axis. Solid (dashed) blue lines show the best fit model (with $\pm 1\sigma$) to the galaxy data. The postage stamp image from SDSS is shown in the top right along with the debiased vote fractions for smooth ($p_s$) and disc ($p_d$) from Galaxy Zoo 2.} }
\label{one_example}
\end{figure}

As the denominator of Equation~\ref{big} is a normalisation factor, comparison between likelihoods for two different SFH models (i.e., two different combinations of $\theta_k = [t_q, \tau]$) is equivalent to a comparison of the numerators. Calculation of $P(\theta_k|d_k, t^{obs})$  for any $\theta$ is possible given galaxy data from the GZ2 sample. Markov Chain Monte Carlo (MCMC; \citealt{MacKay, Dan, GW10}) provides a robust comparison of the likelihoods between $\theta$ values; here we choose \emph{emcee},\footnote{\url{dan.iel.fm/emcee/}} a Python implementation of an affine invariant ensemble sampler by \cite{Dan}.

This method allows for a more efficient exploration of the parameter space by avoiding those areas with low likelihood. A large number of `walkers' are started at an initial position where the likelihood is calculated; from there they individually `jump' to a new area of parameter space. If the likelihood in this new area is greater (less) than the original position then the `walkers' accept (reject) this change in position. Any new position then influences the direction of the  `jumps' of other walkers.  {\changed This is repeated for the defined number of steps after an initial `burn-in' phase. \emph{emcee} returns the positions of these `walkers', which are analogous to the regions of high probability in the model parameter space.} The model outlined above has been coded using the \emph{Python} programming language into a package named \starpy ~which has been made freely available to download\footnote{\url{github.com/zooniverse/starpy}}. {\changed An example output from this Python package for a single galaxy from the GZ2 sample in the red sequence is shown in Figure~\ref{one_example}. The contours show the positions of the `walkers' in the Markov Chain which are analogous to the areas of high probability.}

{\changed We wish to consider the model parameters for the populations of galaxies across the colour magnitude diagram for both smooth and disc galaxies, therefore we run the \starpy ~package on each galaxy in the GZ2 sample. This was extremely time consuming; for each combination of $\theta$ values which \emph{emcee} proposes, a new SFH must be built, prior to convolving it with the BC03 SPS models at the observed age and then predicted colours calculated from the resultant SED. For a single galaxy this takes up to 2 hours on a typical desktop machine for long Markov Chains. A look-up table was therefore generated at $50 ~t^{obs}$, for $100 ~t_{quench}$ and $100 ~\tau$ values; this was then interpolated over for a given observed galaxy's age and proposed $\theta$ values at each step in the Markov Chain. This ensures that a single galaxy takes approximately 2 minutes to run on a typical desktop machine. This interpolation was found to incorporate an error of $\pm 0.04$ into the median $\theta$ values found {\refchange(the 50th percentile position of the walkers}; see Appendix section~\ref{app_lookup} for further information). 

Using this lookup table, each of the $126,316$ total galaxies in the GZ2 sample was run through \starpy ~on multiple cores of a computer cluster to obtain the Markov Chain positions (analogous to $P(\theta_k|d_k)$) for each galaxy, $k$ (see Figure~\ref{one_example}). In each case the Markov Chain consisted of $100$ `walkers' which took $400$ steps in the `burn-in' phase and $400$ steps thereafter, at which point the MCMC acceptance fraction was checked to be within the range $0.25 < f_{acc} < 0.5$ (which was true in all cases). {\newchange Due to the Bayesian nature of this method, a statistical test on the results is not possible; the output is probabilistic in nature across the entirety of the parameter space.} }

\begin{figure*}
\centering{
\includegraphics[width=0.9\textwidth]{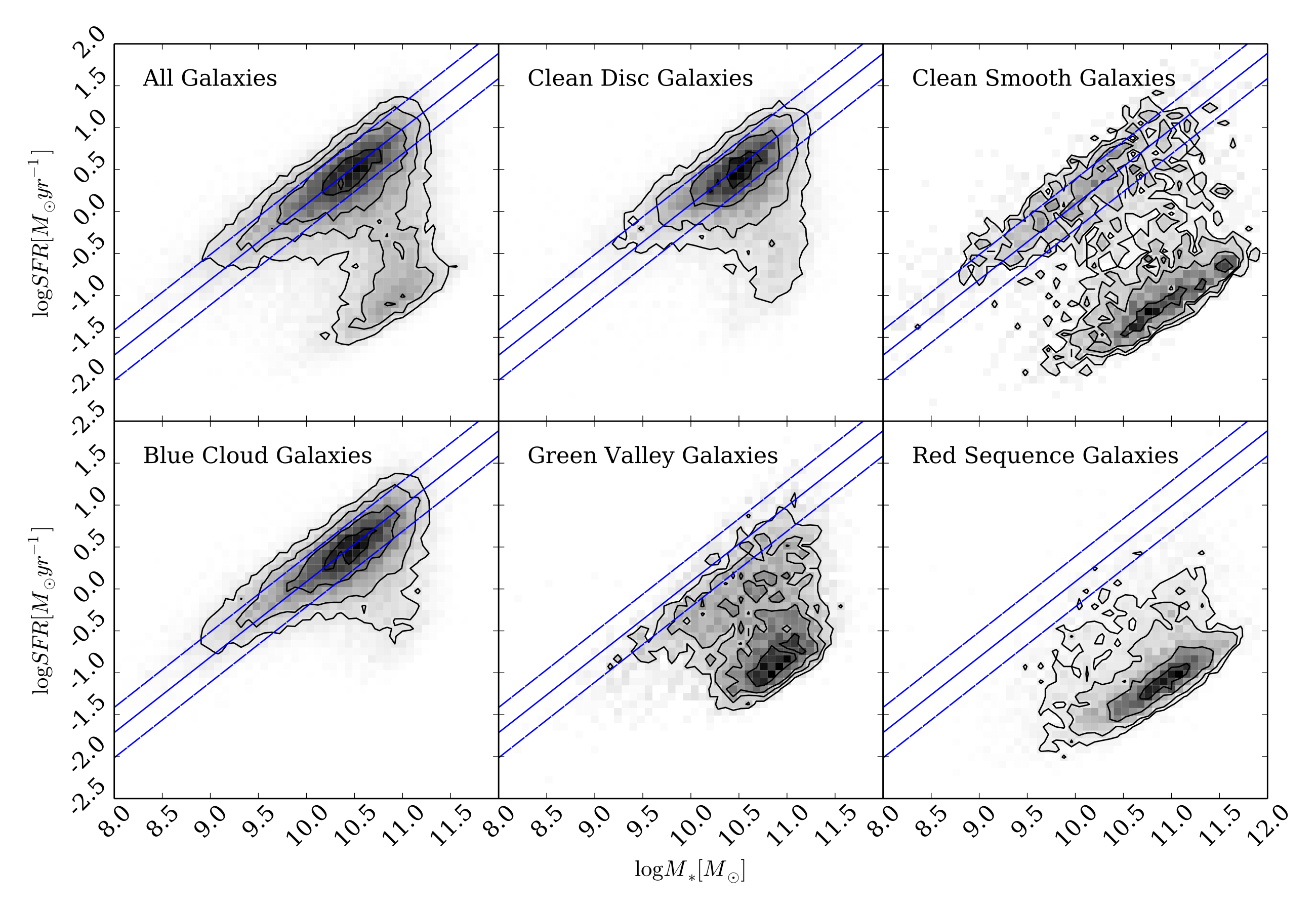}}
\caption{Star formation rate versus stellar mass diagrams show the different populations of galaxies  (top row, left to right: all galaxies, GZ2 `clean' disc and smooth galaxies; bottom row, left to right: blue cloud, green valley and red sequence galaxies) and how they contribute to the star forming sequence (from \citet{Peng}, shown by the solid blue line with 0.3 dex scatter by the dashed lines). Based on positions in these diagrams, the green valley does appear to be a transitional population between the blue cloud and the red sequence. Detailed analysis of star formation histories can elucidate the nature of the different populations' pathways through the green valley. The clean smooth and disc samples are described in Section~\ref{class}.}
\label{sfr_mass_sub}
\end{figure*}

\begin{figure*}
\centering{
\includegraphics[width=\textwidth]{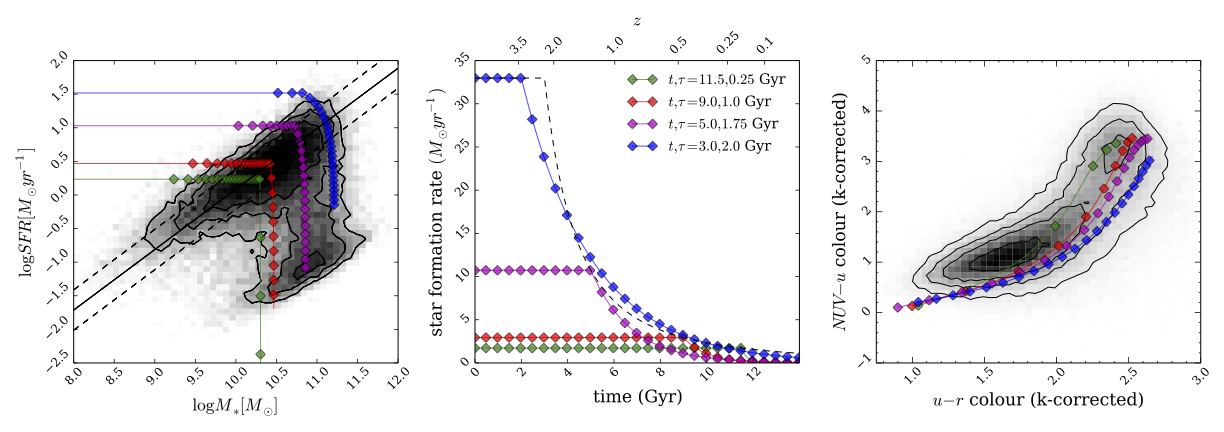}}
\caption{Left panel: SFR vs. $M_*$ for all 126,316 galaxies in our full sample (shaded contours), with model galaxy trajectories shown as coloured points/lines with each point representing a time step of $0.5~\rm{Gyr}$. The SFHs of the models are shown in the middle panel, where the SFR is initially constant before quenching at time $t$ and thereafter exponentially declining with a characteristic timescale $\tau$. We set the SFR at the point of quenching to be consistent with the typical SFR of a star-forming galaxy at the quenching time, $t$ (dashed line; \citealt{Peng}). The full range of models reproduces the observed colour-colour properties of the sample (right panel); for clarity the figures show only 4 of the possible models explored in this study. {\refchange Note that some of the model tracks produce colours redder than the apparent peak of the red sequence in the GZ2 subsample; however this is not the \emph{true} peak of the red sequence due to the necessity for NUV colours from GALEX (see Section \ref{class}).}}
\label{sfr_mass_col}
\end{figure*}

{\newchange These individual galaxy positions are then combined to visualise the areas of high probability in the model parameter space across a given population (e.g. the green valley).} We do this by first discarding positions with a corresponding probability of $P(\theta_k|d_k) < 0.2$ in order to exclude galaxies which are not well fit by the quenching model; for example blue cloud galaxies which are still star forming will be poorly fit by a quenching model (see Section~\ref{qmod}). {\refchange Using this constraint, $2.4\%$, $7.0\%$ and $5.4\%$ of green, red and blue galaxies respectively had \emph{all} of their walker positions discarded. These are not significant enough fractions to affect the results (see Appendix section \ref{discard} for more information.)} The Markov Chain positions are then binned and weighted by their {\newchange corresponding logarithmic posterior probability $\log [P(\theta_k|d_k)]$, provided by the \emph{emcee} package, to further emphasise the features and differences between each population in the visualisation}. The GZ2 data also provides uniquely powerful continuous measurements of a galaxy's morphology, therefore we utilise the user vote fractions to obtain separate model parameter distributions for both smooth and disc galaxies. This is obtained by also weighting by the morphology vote fraction when the binned positions are summed. {\newchange We stress that this portion of the methodology is a non-Bayesian visualisation of the combined individual galaxy results for each population.}

For example, the galaxy shown in Figure~\ref{one_example} would contribute almost evenly to both the smooth and disc parameters due to the GZ2 vote fractions. Since galaxies with similar vote fractions contain both a bulge and disc component, this method is effective in incorporating intermediate galaxies which are thought to be crucial to the morphological changes between early- and late-type galaxies. It was the consideration of these intermediate galaxies which was excluded from the investigation by S14.

\section{Results}\label{results}
\subsection{Initial Results}

\begin{figure*}
\includegraphics[width=0.4975\textwidth]{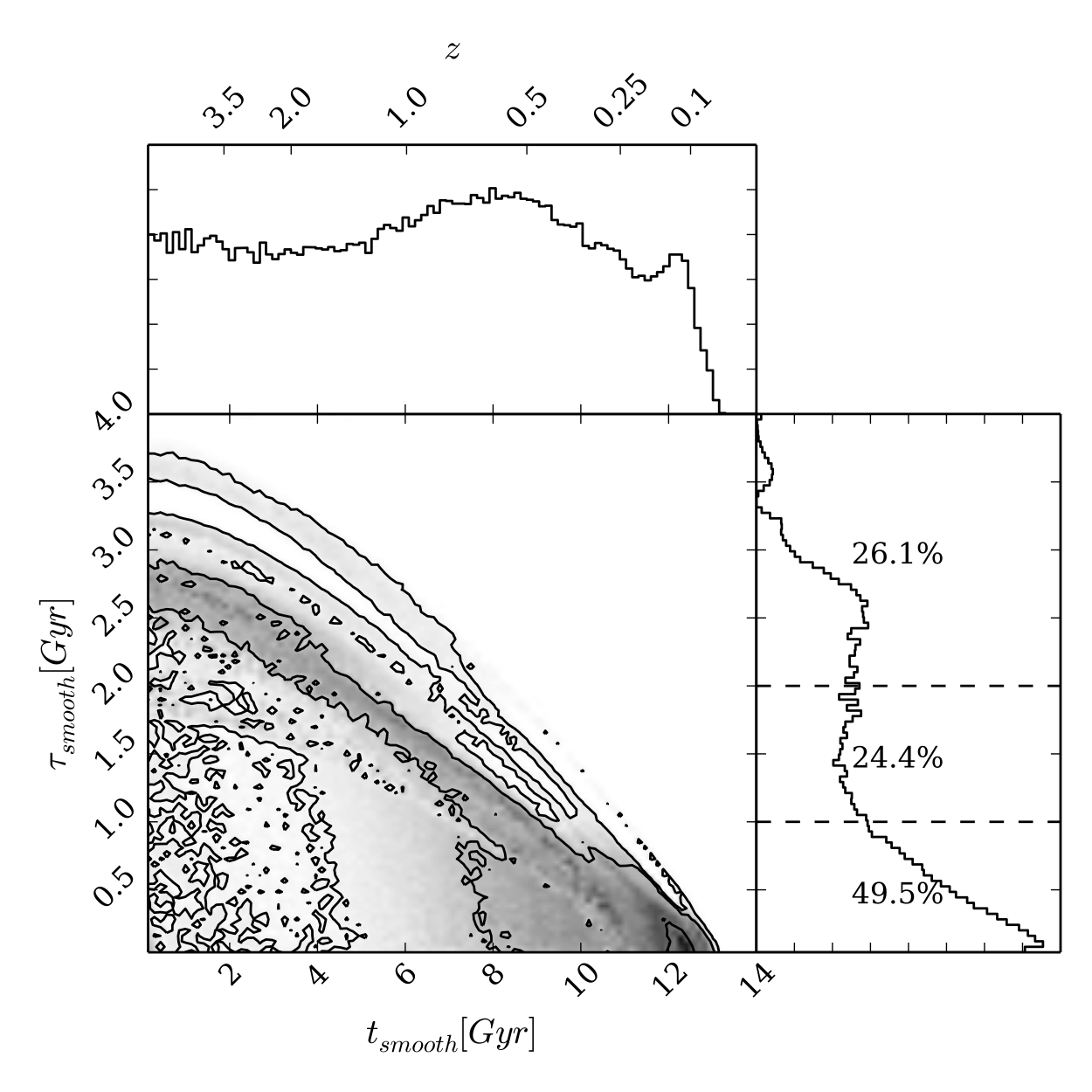}
\includegraphics[width=0.4975\textwidth]{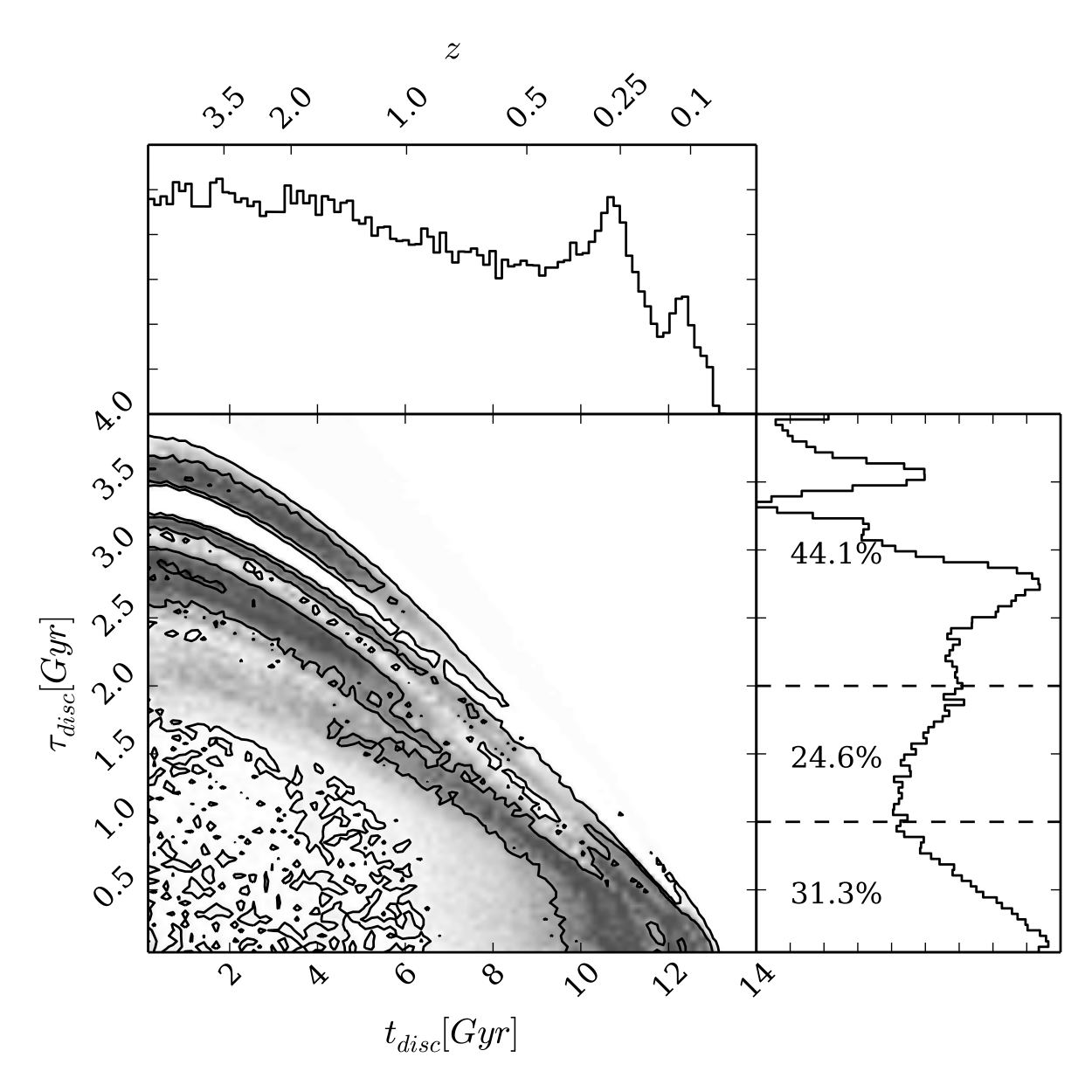}
\caption[8pt]{{\changed Contour plots showing the combined positions in the Markov Chain for all red galaxies in this study, weighted by the logarithmic probability of each position (see Section \ref{stats}) and also by the morphological vote fractions from GZ2 to give the areas of high probability in the model parameter space for both bulge (left) and disc (right) dominated systems. The histograms show the projection into one dimension for each parameter. The dashed lines show the separation between rapid ($\tau ~\rm{[Gyr]} < 1.0$), intermediate ($1.0 < \tau ~\rm{[Gyr]} < 2.0$) and slow ($\tau ~\rm{[Gyr]} > 2.0$) quenching timescales with the fraction of the combined posterior probability distribution in each region shown (see Section~\ref{stats}).}}
\label{red_s}
\end{figure*}

Figure~\ref{sfr_mass_sub} shows the SFR versus the stellar mass for the observed GZ2 sample which has been split into blue cloud, green valley and red sequence populations as well as into the `clean' disc and smooth galaxy samples (with GZ2 vote fractions of $p_d \geq 0.8$ and $p_s \geq 0.8$ respectively). The green valley galaxies are indeed a population which have either left, or begun to leave, the star forming sequence or have some residual star formation still occurring.

The left panel in Figure~\ref{sfr_mass_col} shows a handful of quenching models and how they reproduce the observed relationship between the SFR and the mass of a galaxy, including how at the time of quenching they reside on the star forming sequence shown by the solid black line for a galaxy of mass, $M = 10^{10.27} M_{\odot}$.  {\refchange The right panel shows how these SFRs translate into the optical-NUV colour-colour plane to reproduce observed colours of green valley and red sequence galaxies. Some of the SFHs produce colours redder than the apparent peak of the red sequence in the GZ2 subsample; however this is not the \emph{true} peak of the red sequence due to the necessity for NUV colours from GALEX (see Section \ref{class}). 

The majority of the red galaxies in the sample therefore lie towards the \emph{blue end} of the red sequence and have a small amount of residual star formation in order to be detected in the NUV {\tworef resulting in a specific subset of the red sequence studied in this investigation. Only $47\%$ of the red sequence galaxies present in the entire Galaxy Zoo 2 sample are matched with GALEX to produce our final sample of $126, 316$ galaxies, as opposed to $72\%$ of the blue cloud and $53\%$ of the green valley galaxies.} This limitation should be taken into account when considering the results in the following sections.}

{\changed The SFH models were implemented with the \starpy ~package to produce Figures~\ref{red_s},~\ref{green_v} \&~\ref{blue_c} for the red sequence, green valley and blue cloud populations of smooth and disc galaxies respectively.} {\newchange The percentages shown in Figures~\ref{red_s},~\ref{green_v} \&~\ref{blue_c} are calculated as the fractions of the combined posterior probability distribution located in each region of parameter space for a given population. 

Since the sample contains such a large number of galaxies, we interpret these fractions as broadly equivalent to the percentage of galaxies in a given population undergoing quenching within the stated timescale range. Although this is not quantitatively exact, it is nevertheless a useful framework for interpreting the results of combining the individual posterior probability distributions of each galaxy.} 

{\refchange Also shown in Figure~11 are the median walker positions (the 50th percentile of the Bayesian probability distribution) of each individual galaxy, split into red, green and blue populations also with a hard cut in the vote fraction of $p_d > 0.5$ and $p_s > 0.5$ to show the disc and smooth populations respectively. These positions were calculated without discarding any walker positions due to low probability and without weighting by vote fractions; therefore this may be more intuitive to understand than Figures~\ref{red_s},~\ref{green_v} \&~\ref{blue_c}.}

{\refchange Although the quenching timescales are continuous in nature, in this Section we refer to rapid, intermediate and slow quenching timescales which correspond to ranges of {\changed $\tau ~\rm{[Gyr]} < 1.0$, $1.0 < \tau ~\rm{[Gyr]} < 2.0$ and $\tau ~\rm{[Gyr]} > 2.0$} respectively for ease of discussion.}

\subsection{The Red Sample}\label{rs}

The left panel of Figure~\ref{red_s} reveals that {\tworef smooth galaxies with red optical colours} {\changed show a preference $(49.5\%$; see Figure~\ref{red_s})} for rapid quenching timescales across all cosmic time resulting in a very low current SFR. 
{\changed For these smooth red galaxies we see, at early times only, a preference for slow and intermediate timescales in the left panel of Figure~\ref{red_s}. Perhaps this is the influence of intermediate galaxies (with $p_s \sim p_d \sim 0.5$), hence why similar high probability areas exist for both the smooth-like and disc-like galaxies in the left and right panels of Figure~\ref{red_s}}. This is especially apparent considering there are far more of these intermediate galaxies than those that are definitively early- or late-types (see Table~\ref{subs}). These galaxies are those whose morphology cannot be easily distinguished either because they are at a large distance or because they are an S0 galaxy whose morphology can be interpreted by different users in different ways. \citet{GZ2} find that S0 galaxies expertly classified by \citet{NA10} are more commonly classified as ellipticals by GZ2 users, but have a significant tail to high disc vote fractions, giving a possible explanation as to the origin of this area of probability.

{\changed The right panel of Figure~\ref{red_s} reveals that red disc galaxies show similar preferences for rapid $(31.3\%)$ and slow $(44.1\%)$ quenching timescales. The preference for \emph{very} slow ($\tau > 3.0 ~\rm{Gyr}$) quenching timescales (which are not seen in either the green valley or blue cloud, see Figures~\ref{green_v} and~\ref{blue_c})} suggests that these  galaxies have only just reached the red sequence after a very slow evolution across the colour-magnitude diagram. Considering their limited number and our requirement for NUV emission, it is likely that these galaxies are currently on the edge of the red sequence having recently (and finally) moved out of the green valley. Table~\ref{subs} shows that $3.9\%$ of our sample are red sequence clean disc galaxies, i.e. red late-type spirals. This is, within uncertainties, in agreement with the findings of \citet{Masters10}, who find $\sim6\%$ of late-type spirals are red when defined by a cut in the $g-r$ optical colour (rather than with $u-r$ as implemented in this investigation) and are at the `blue end of the red sequence'. 

{\newchange Despite the dominance of slow quenching timescales, the red disc galaxies also show some preference for rapid quenching timescales ($31.3\%$), similar to the red smooth galaxies but with a lower probability. Perhaps these rapid quenching timescales can also be attributed to a morphological change, suggesting that the quenching has occurred more rapidly than the morphological change to a bulge dominated system.}

Comparing the resultant SFRs for both the smooth- and disc-like galaxies in Figure~\ref{red_s} by noticing {\changed where the areas of high probability lie with respect to the bottom panel of Figure~\ref{pred} (which shows the predicted SFR at an observation time of $t\sim12.8~\rm{Gyr}$, the average `observed' time of the GZ2 population)} reveals that red disc galaxies with a preference for slow quenching still have some residual star formation occurring, SFR$~\sim0.105 M_{\odot}yr^{-1}$, whereas the smooth galaxies with a dominant preference for rapid quenching have a resultant SFR$~\sim0.0075 M_{\odot}yr^{-1}$. This is approximately 14 times less than the residual SFR still occurring in the red sequence disc galaxies. Within error, this is in agreement with the findings of \citet{Toj13} who, by using the VErsatile SPectral Analyses spectral fitting code {\newchange (VESPA; \citealt{Tojero07})}, found that red late-type spirals show 17 times more recent star formation than red elliptical galaxies.

These results for the red galaxies investigated here with NUV emission, have many implications for green valley galaxies, as all of these systems must have passed through the green valley on their way to the red sequence. 

\subsection{Green Valley Galaxies}\label{gv}

\begin{figure*}
\includegraphics[width=0.4975\textwidth]{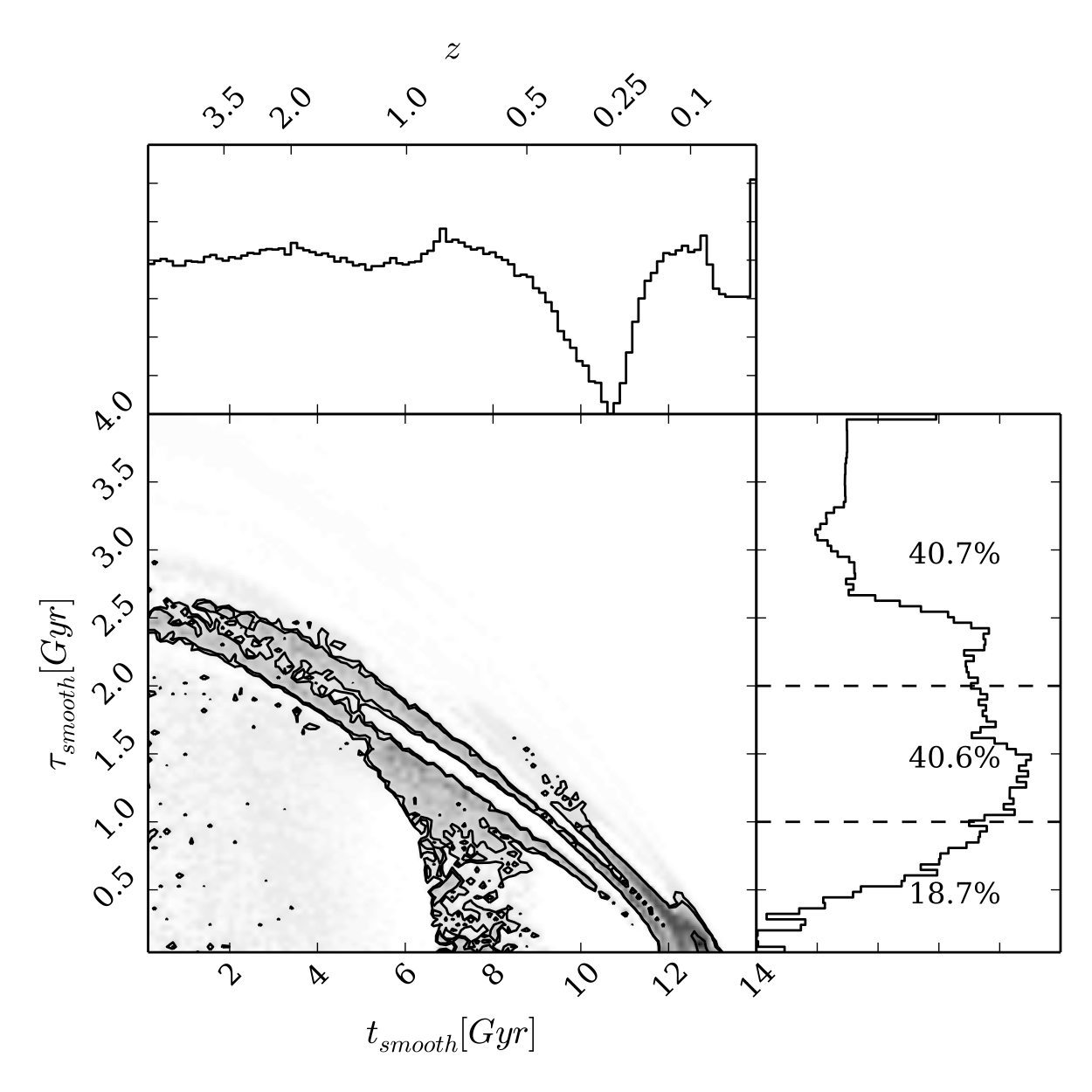}
\includegraphics[width=0.4975\textwidth]{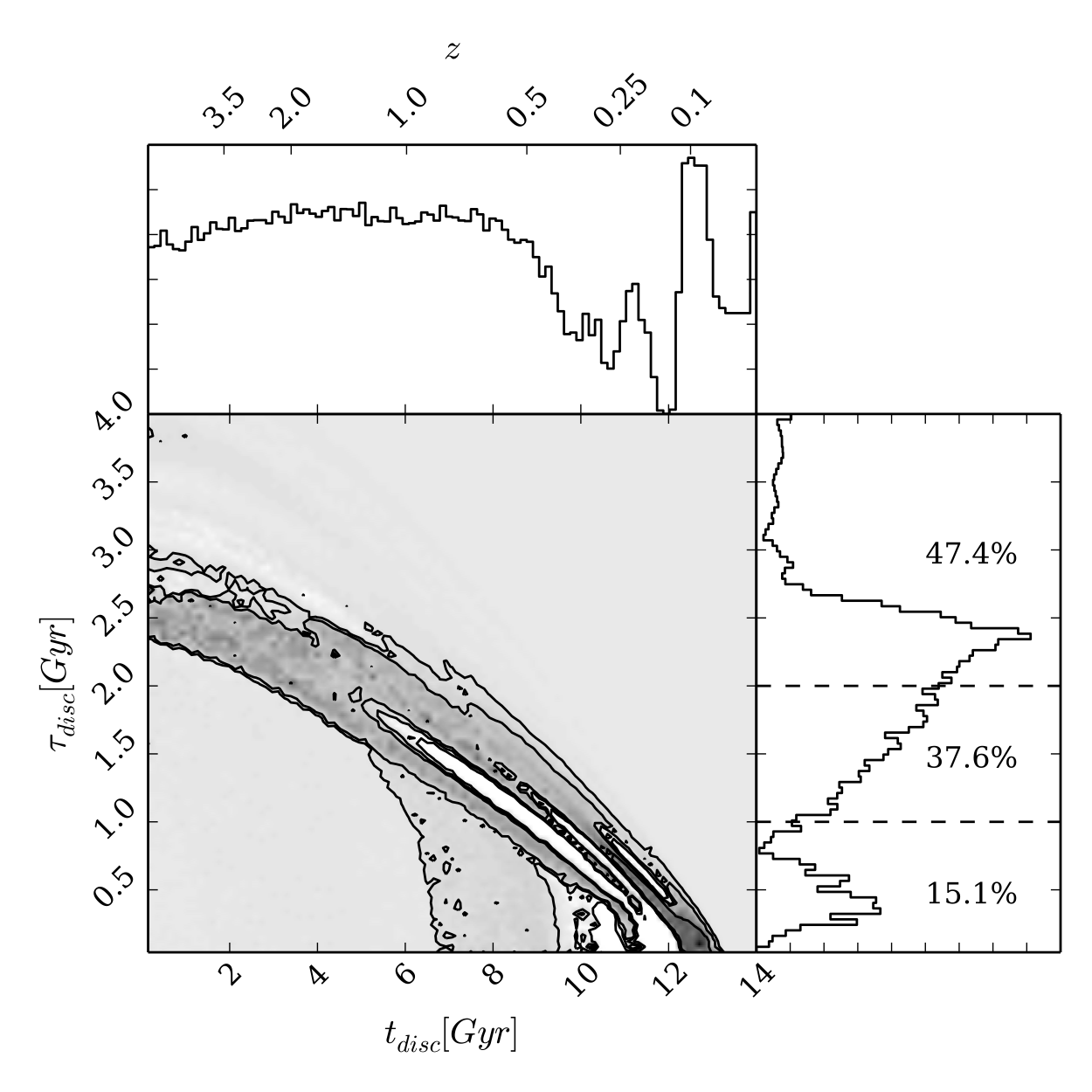}
\caption[8pt]{{\changed Contour plots showing the combined positions in the Markov Chain for galaxies in the green valley, weighted by the logarithmic probability of each position (see Section \ref{stats}) and also by the morphological vote fractions from GZ2 to give the areas of high probability in the model parameter space for both bulge (left) and disc (right) dominated systems. The histograms show the projection into one dimension for each parameter. The dashed lines show the separation between rapid ($\tau ~\rm{[Gyr]} < 1.0$), intermediate ($1.0 < \tau ~\rm{[Gyr]} < 2.0$) and slow ($\tau ~\rm{[Gyr]} > 2.0$) quenching timescales with the fraction of the combined posterior probability distribution in each region shown (see Section~\ref{stats}).}}
\label{green_v}
\end{figure*}

In Figure~\ref{green_v} we can make similar comparisons for the green valley galaxies to those discussed previously for the subset of red galaxies studied. {\newchange For the red galaxies, an argument can be made for two possible tracks across the green valley, shown by the bimodal nature of both distributions in $\tau$ with a common area in the intermediate timescales region where the rapid and slow timescales peaked distributions intersect. However in the green valley this intermediate quenching timescale region becomes more significant {\refchange (in agreement with the conclusions of \citealt{Gonc12})}, particularly for the smooth-like galaxies (see the left panel of Figure~\ref{green_v}). }

The smooth galaxy parameters favour these intermediate quenching timescales ($40.6\%$) with some preference for slow quenching at  early times ($z > 1$). The preference for rapid quenching of smooth galaxies has dropped by over a half compared to the red galaxies, {\refchange however this will be influenced by the observability of galaxies undergoing such a rapid quench which will spend significantly less time in the transitional population of the green valley}. {\changed Those galaxies with such a rapid decline in star formation will pass so quickly through the green valley they will be detected at a lower number than those galaxies which have stalled in the green valley with intermediate quenching timescales;} accounting for the observed number of intermediate galaxies which are present in the green valley {\newchange and the dominance of rapid timescales detected for red galaxies for both morphologies.} 

{\changed The disc galaxies of the green valley now overwhelmingly prefer slow quenching timescales ($47.4\%$) with a similar amount of intermediate quenching compared to the smooth galaxy parameters ($37.6\%$; see Figure~\ref{green_v}).} There is still some preference for galaxies with a star formation history which results in a high current SFR, suggesting there are also some late-type galaxies that have just progressed from the blue cloud into the green valley. 

{\changed If we compare Figure~\ref{green_v} to Figure~\ref{red_s} {\refchange we can see quenching has occurred at later (more recent) cosmic times} in the green valley {\tworef at least for red galaxies} for both morphological types.} Therefore both morphologies are tracing the evolution of the red sequence, confirming that the green valley is indeed a transitional population between blue cloud and red sequence regardless of morphology. Currently as we observe the green valley, its main constituents are very slowly evolving disc-like galaxies along with intermediate- and smooth-like galaxies which pass across it with intermediate timescales within $\sim 1.0-1.5~\rm{Gyr}$.

Given enough time ($t\sim4 - 5~\rm{Gyr}$), the disc galaxies will eventually fully pass through the green valley and make it out to the red sequence (the right panel of Figure~\ref{sfr_mass_col} shows galaxies with $\tau > 1.0~\rm{Gyr}$ do not approach the red sequence within $3~\rm{Gyr}$ post quench). This is most likely the origin of the `red spirals'.

\begin{figure*}
\includegraphics[width=0.4975\textwidth]{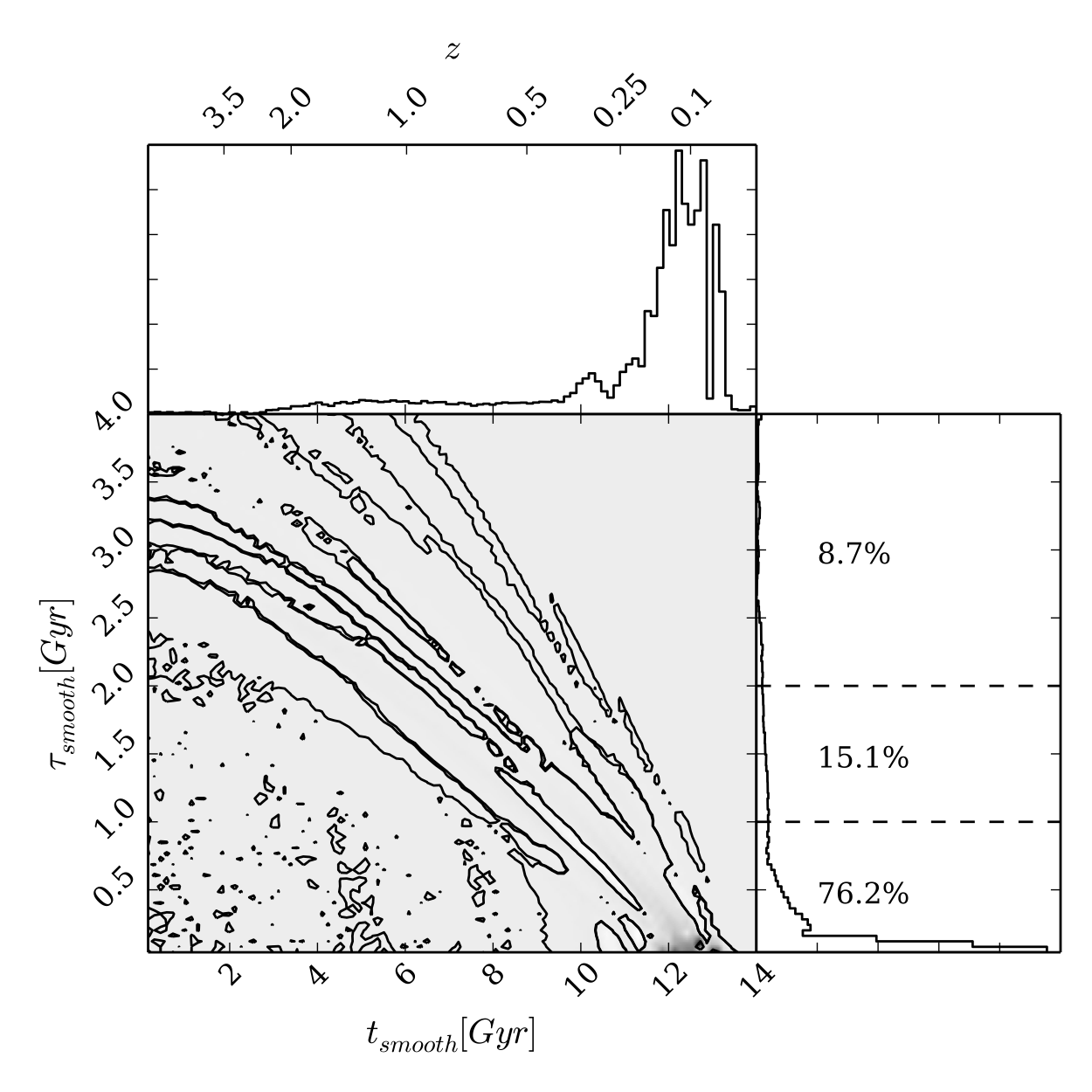}
\includegraphics[width=0.4975\textwidth]{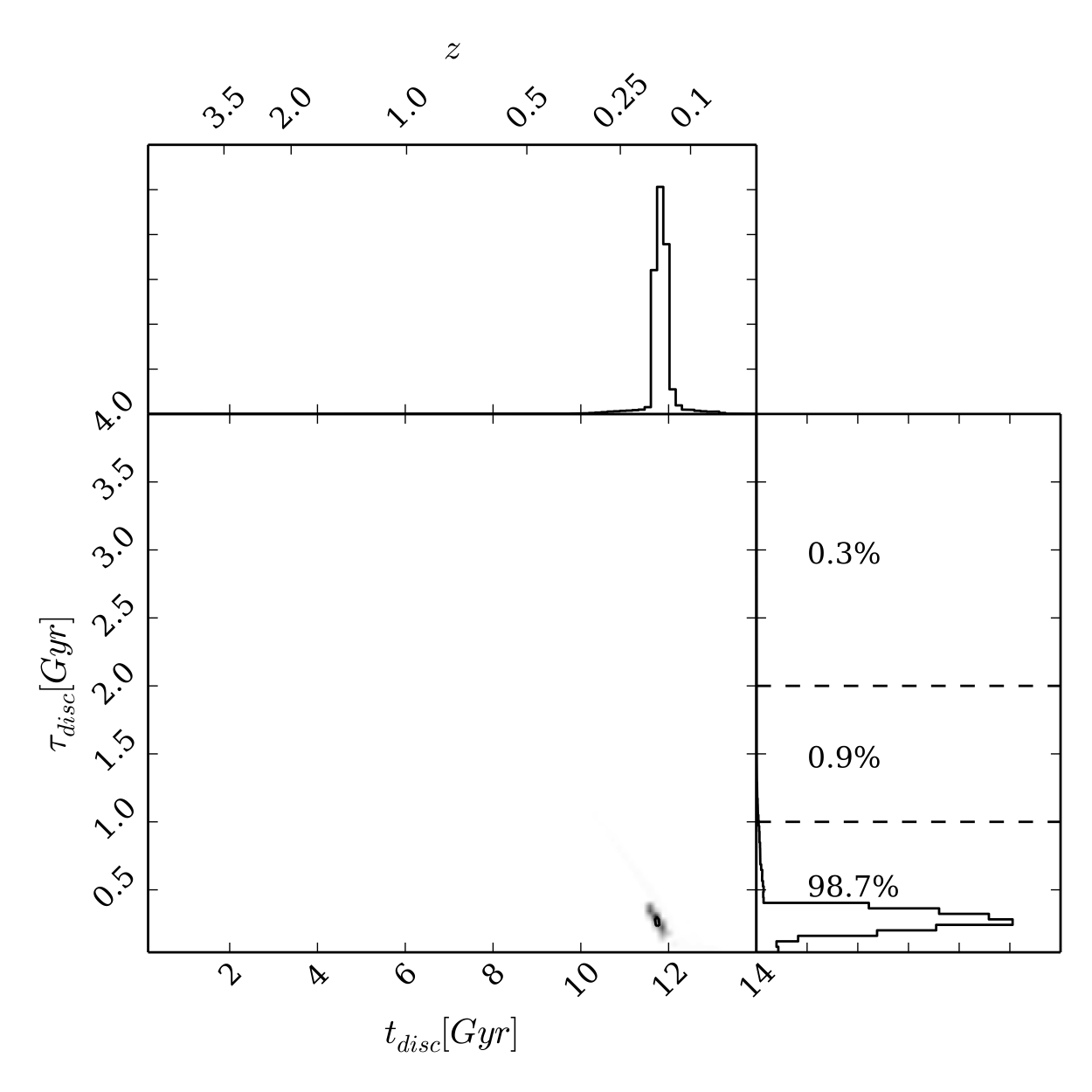}
\caption[8pt]{{\changed Contour plots showing the combined positions in the Markov Chain for galaxies in the blue cloud, weighted by the logarithmic probability of each position (see Section \ref{stats}) and also by the morphological vote fractions from GZ2 to give the areas of high probability in the model parameter space for both bulge (left) and disc (right) dominated systems. The histograms show the projection into one dimension for each parameter. The dashed lines show the separation between rapid ($\tau ~\rm{[Gyr]} < 1.0$), intermediate ($1.0 < \tau ~\rm{[Gyr]} < 2.0$) and slow ($\tau ~\rm{[Gyr]} > 2.0$) quenching timescales with the fraction of the combined posterior probability distribution in each region shown (see Section~\ref{stats}). Positions with probabilities less than 0.2 are discarded as poorly fit models, therefore we can conclude unsurprisingly that blue cloud galaxies are not well described by a quenching star formation model. }}
\label{blue_c}
\end{figure*}

{\changed If we consider {\tworef then that the green valley is a transitional population}, then we can expect that the ratio of smooth:disc galaxies that is currently observed in the green valley will evolve into the ratio observed for {\tworef the red galaxies with NUV emission investigated}. Table~\ref{subs} shows the ratio of smooth : disc galaxies in the observed red sequence of the GZ2 sample is $62:38$ whereas in the green valley it is $45:55$. {\refchange Making the very simple assumptions that this ratio does not change with redshift and that quenching is the only mechanism which causes a morphological transformation, we can infer that $31.2\%$ of the disc-dominated galaxies currently residing in the green valley would have to undergo a morphological change to a bulge-dominated galaxy.} We find that the fraction of the probability for green valley disc galaxies occupying the parameter space $\tau < 1.5 ~\rm{Gyr}$ is $29.4\%$, and therefore suggest that quenching mechanisms with these timescales are capable of destroying the disc-dominated nature of galaxies. {\refchange This is most likely an overestimate of the mechanisms with timescales that can cause a morphological change because of the observability of those galaxies which undergo such a rapid quench; \citet{Martin07} showed that after considering the time spent in the green valley, the fraction of galaxies undergoing a rapid quench quadruples.}}

All of this evidence suggests that there are not just two routes for galaxies through the green valley {\refchange  as concluded by S14}, but a continuum of quenching timescales which we can divide into three general regimes: {\newchange rapid ($\tau < 1.0 ~\rm{Gyr}$), intermediate ($1.0 < \tau < 2.0~\rm{Gyr}$) and slow ($\tau > 2.0~\rm{Gyr}$). The intermediate quenching timescales reside in the space between the extremes sampled by the UV/optical diagrams of S14; the inclusion of the intermediate galaxies in this investigation (unlike in S14) and the more precise Bayesian analysis, quantifies this range of $\tau$ and specifically ties the intermediate timescales to all variations of galaxy morphology.}

\subsection{Blue Cloud Galaxies}\label{bc}

\begin{figure*}\label{bestfit}
\includegraphics[width=0.95\textwidth]{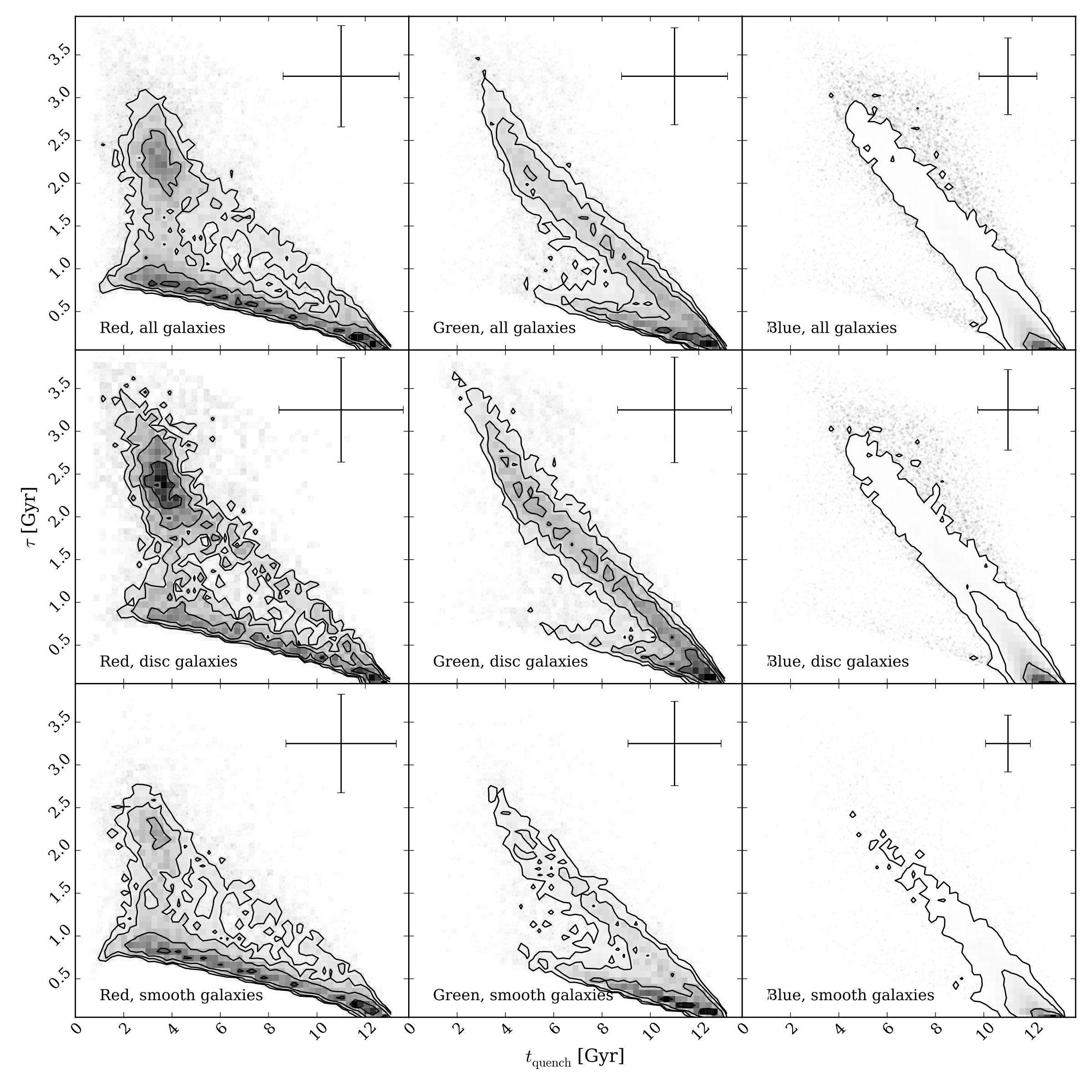}
\caption{Contours showing the positions in the $[t, \tau]$ parameter space of the median walker position (the 50th percentile; as shown by the intersection of the solid blue lines in Figure~\ref{one_example}) for each galaxy for all (top), disc ($p_d > 0.5$; middle), and smooth ($p_s > 0.5$; bottom) red sequence, green valley and blue cloud galaxies in the left, middle and right panels respectively. The error bars on each panel shows the average $68\%$ confidence on the median positions {\refchange (calculated from the 16th and 84th percentile,} as shown by the blue dashed lines in Figure~\ref{one_example}). These positions were calculated without discarding any walker positions due to low probability and without weighting by vote fractions, therefore this plot may be more intuitive than Figures~\ref{red_s},~\ref{green_v} \&~\ref{blue_c}. The differences between the smooth and disc populations and between the red, green and blue populations remain clearly apparent.}
\end{figure*}

Since the blue cloud is considered to be primarily made of star forming galaxies we expect \starpy~ to have some difficulty in determining the most likely quenching model to describe them, as confirmed by Figure~\ref{blue_c}. The attempt to characterise a star forming galaxy with a quenched SFH model leads \starpy~ to attribute the extremely blue colours of the majority of these galaxies to fast quenching at recent times (i.e. very little change in the SFR; see the right panel of Figure~\ref{blue_c} in comparison with the bottom panel of Figure~\ref{pred}).

{\changed This is particularly apparent for the blue disc population. Perhaps even galaxies which are currently quenching slowly across the blue cloud cannot be well fit by the quenching models implemented, as they still have high SFRs despite some quenching (although a galaxy has undergone quenching, star formation can still occur in a galaxy, just at a slower rate than at earlier times, described by $\tau$).}

There is a very small preference among blue bulge dominated galaxies for slow quenching which began prior to $z \sim 0.5 $. These populations have been blue for a considerable period of time, slowly using up their gas for star formation by the Kennicutt$-$Schmidt law \citep{Schmidt59, Kennicutt97}. However the major preference is for rapid quenching at recent times in the blue cloud; this therefore provides some support to the theories for blue ellipticals as either merger-driven {\refchange ($\sim76\%$; like those identified as recently quenched ellipticals with properties consistent with a merger origin by \citealt{McIntosh14}) or gas inflow-driven reinvigorated star formation that is now slowly decreasing ($\sim24\%$; such as the population of blue spheroidal galaxies studied by \citealt{Kaviraj13}).} However, we remind the reader that the quenching models used in this work do not provide an adequate fit to the blue cloud population.

The blue cloud is therefore primarily composed of both star forming galaxies with any morphology and smooth galaxies which {\changed are undergoing a rapid quench, presumably after a previous event triggered star formation and turned them blue}.

\section{Discussion}\label{diss}

We have implemented a Bayesian statistical analysis of the star formation histories (SFHs) of a large sample of galaxies morphologically classified by Galaxy Zoo. We have found differences between the SFHs of smooth- and disc-like galaxies across the colour-magnitude diagram in the red sequence, green valley and blue cloud. In this section we will speculate on the question: what are the possible mechanisms driving these differences? 

\subsection{Rapid Quenching Mechanisms}\label{rapid}

{\changed Rapid quenching is much more prevalent in smooth galaxies than disc galaxies, and the {\tworef red galaxies with NUV emission in this study} are also much more likely to be characterised by a rapid quenching model than green valley galaxies (ignoring blue cloud galaxies due to their apparent poor fit by the quenching models, see Figure~\ref{blue_c}). In the green valley there is also a distinct lack of preference for rapid quenching timescales with $\tau < 0.5~\rm{Gyr}$; {\refchange however we must bear in mind the observability of a rapid quenching history declines with decreasing $\tau$. Rapid mechanisms may be more common in the green valley than seen in Figure \ref{green_v}, however this observability should not depend on morphology so we can still conclude that rapid quenching mechanisms are detected more for smooth rather than disc galaxies.}} This suggests that this rapid quenching mechanism causes a change in morphology from a disc- to a smooth-like galaxy as it quickly traverses the colour-magnitude diagram to the red sequence, {\newchange supported by the number of disc galaxies that would need to undergo a morphological change in order for the disc : smooth ratio of galaxies in the green valley to match that of the red galaxies (see Section~\ref{gv})}. {\newchange From this indirect evidence we suggest that} this rapid quenching mechanism is due to major mergers.

Inspection of the galaxies contributing to this area of probability reveals that this does not arise due to \emph{currently} merging pairs missed by GZ users which were therefore not excluded from the sample (see Section \ref{class}), but by typical smooth galaxies with red optical and NUV colours that the model attributes to rapid quenching at early times. {\refchange Although a prescription for modelling a merger in the SFH is not included in this work we can still detect the after effects (see Section \ref{future} for future work planned with \starpy).} 

One simulation of interest by \citet{Springel05} showed that feedback from black hole activity is a necessary component of destructive major mergers to produce such rapid quenching timescales. Powerful quasar outflows remove much of the gas from the inner regions of the galaxy, terminating star formation on extremely short timescales. \citet{Bell06}, using data from the COMBO-17 redshift survey ($0.4 < z < 0.8$), estimate a merger timescale from being classified as a close galaxy pair to recognisably disturbed as $\sim 0.4~\rm{Gyr}$. \citet{Springel05} consequently find using hydrodynamical simulations that after $\sim1~\rm{Gyr}$ the merger remnant has reddened to $u-r \sim 2.0$. This is in agreement with our simple quenching models which show (Figure~\ref{sfr_mass_col}) that within $\sim1~\rm{Gyr}$ the models with a SFH with $\tau < 0.4~\rm{Gyr}$ have reached the red sequence with $u-r ~\ga 2.2$. {\changed This could explain the preference for red disc galaxies with rapid quenching timescales ($31.3\%$), as they may have undergone a major merger recently but are still undergoing a morphological change from disc, to disturbed, to an eventual smooth galaxy (see also \citealt{vdW09}).} 

{\changed We reiterate that this rapid quenching mechanism occurs much more rarely in green valley galaxies of both morphologies than {\tworef for the subset of red sequence galaxies studied}, however does not fully characterise all the galaxies in either the red sequence or green valley.} Dry major mergers therefore do not fully account for the formation of any galaxy type at any redshift, supporting the observational conclusions made by \citet{Bell07,Bundy07, Kav14a} and simulations by \citet{Genel08}. 

\subsection{Intermediate Quenching Mechanisms}\label{int}
{\changed Intermediate quenching timescales are found to be equally prevalent across populations for both smooth and disc galaxies across cosmic time, {\newchange particularly in the green valley. Intermediate timescales are the prevalent mechanism for quenching smooth green valley galaxies, unlike the rapid quenching prevalent for red galaxies.}} We suggest that this intermediate quenching route must therefore be possible with routes that both preserve and transform morphology. It is this result of {\newchange another route through the green valley that is in contradiction} with the findings of S14. 

If we once again consider the simulations of \citet{Springel05}, this time without any feedback from black holes, they suggest that if even a small fraction of gas is not consumed in the starburst following a merger (either because the mass ratio is not large enough or from the lack of strong black hole activity) the remnant can sustain star formation for periods of several Gyrs. The remnants from these simulations take $\sim5.5~\rm{Gyr}$ to reach red optical colours of $u-r \sim 2.1$. We can see from Figure~\ref{sfr_mass_col} that the models with intermediate quenching timescales of $1.0 \la ~\tau~\rm{[Gyr]} ~\la 2.0$ take approximately $2.5-5.5~\rm{Gyr}$ to reach these red colours.

We speculate that the intermediate quenching timescales are caused by gas rich major mergers, major mergers without black hole feedback and from minor mergers, the latter of which is the dominant mechanism. This is supported by the findings of \citet{Lotz08} who find that the detectability timescales for equal mass gas rich mergers with large initial separations range from $\sim 1.1-1.9~\rm{Gyr}$, and of \citet{Lotz11}, who find in further simulations that as the baryonic gas fraction in a merger with mass ratios of 1:1-1:4 increases, so does the timescale of the merger from $\sim0.2~\rm{Gyr}$ (with little gas, as above for major mergers causing rapid quenching timescales) up to $\sim1.5~\rm{Gyr}$ (with large gas fractions). {\refchange Here we are assuming that the morphologically detectable timescale of a merger is roughly the same order as the quenching timescale. However, we must consider the existence of a substantial population of blue ellipticals \citep{Sch09}, which are thought to be post-merger systems with no detectable morphological signatures of a merger but with the merger-induced starburst still detectable in the photometry. This photometry is an indicator for the SFH and therefore should present with longer timescales for the photometric effects of a merger than found in the simulations by \citet{Lotz08} and \citet{Lotz11}. Observing this link between the timescale for the morphological observability of a merger and the timescales for the star formation induced by a merger is problematic, as evidenced by the lack of literature on the subject.}

\citet{Lotz08} also show that the remnants of these simulated equal mass gas rich disc mergers (wet disc mergers) are observable for $\ga1~\rm{Gyr}$ post merger and {\refchange state that they appear ``disc-like and dusty" in the simulations, which is consistent with an ``early-type spiral morphology"}.  Such galaxies are often observed to have spiral features with a dominant bulge, suggesting that such galaxies may divide the votes of the GZ2 users, producing vote fractions of $p_s \sim p_d \sim 0.5$. We believe this is why the intermediate quenching timescales are equally dominant for both smooth and disc galaxies across each population in Figures~\ref{red_s} and~\ref{green_v}. 

Other simulations (e.g. such as \citet{Rob06} and \citet{Barnes02}) support the conclusion that both gas rich major mergers and minor mergers can produce disc-like remnants. Observationally, \citet{Darg10a} showed an increase in the spiral to elliptical ratio for merging galaxies ($0.005 < z < 0.1$) by a factor of two compared to the general population. They attribute this to the much longer timescales during which mergers of spirals are observable compared to mergers with elliptical galaxies, {\changed confirming our hypothesis that the quenching timescales $\tau < 1.5 ~\rm{Gyr}$ preferred by disc galaxies may be undergoing mergers which will eventually lead to a morphological change}. Similarly, \citet{Casteels13} observe that galaxies ($0.01 < z < 0.09$) which are interacting often retain their spiral structures and that a spiral galaxy which has been classified as having `loose winding arms' by the GZ2 users are often entering the early stages of mergers and interactions.

{\changed $40.6\%$ of the probability for smooth galaxies in the green valley arises due to intermediate quenching timescales (see Figure~\ref{green_v}); this is in agreement with work done by \citet{Kav14a, Kav14b} who by studying SDSS photometry ($z<0.07$) state that approximately half of the star formation in galaxies is driven by minor mergers at $0.5 < z < 0.7$ therefore exhausting available gas for star formation and consequently causing a gradual decline in the star formation rate}. This supports earlier work by \cite{Kav11} who, using multi wavelength photometry of galaxies in COSMOS \citep{Scoville07}, found that $70\%$ of early-type galaxies appear morphologically disturbed, suggesting either a minor or major merger in their history. {\changed This is in agreement with the total percentage of probability with $\tau < 2.0 ~\rm[Gyr]$; $73.9\%$ and $59.3\%$, for the smooth red and green galaxies in Figures~\ref{red_s} and~\ref{green_v} respectively.} {\refchange Note that the star formation model used here is a basic one and has no prescription for reignition of star formation post-quench which can also cause morphological disturbance of a galaxy, like those detected by \cite{Kav11}.}

{\refchange \citet{Darg10a} show in their Figure 6 that that beyond a merger ratio of $1:10$ (up to $\sim 1:100$), green is the dominant average galaxy colour of the visually identified merging pair in GZ. These mergers are also dominated by spiral-spiral mergers as opposed to elliptical-elliptical and elliptical-spiral. This supports our hypothesis that these intermediate timescales dominating in the green valley are caused in part by minor mergers. This is contradictory to the findings of \citet{Mendez11} who find the merger fraction in the green valley is much lower than in the blue cloud, however they use an analytical light decomposition indicator ($Gini/M_{20}$; see \citealt{Lotz08}) to identify their mergers, which tend to detect major mergers more easily than minor mergers. We have discussed the lower likelihood of a green valley galaxy to undergo a rapid quench, which we have attributed to major mergers (see Section \ref{rapid}), despite the caveat of the observability and believe that this may have been the phenomenon that \citet{Mendez11} detected.}

The resultant intermediate quenching timescales occur due to one interaction mechanism, unlike the rapid quenching, which occurs due to a major merger combined with AGN feedback, and decreases the SFR over a short period of time. Therefore any external event which can cause either a burst of star formation (depleting the gas available) or directly strip a galaxy of its gas, {\refchange for example galaxy harassment, interactions, ram pressure stripping, strangulation and interactions internal to clusters, would cause quenching on an intermediate timescale. Such mechanisms would be the dominant cause of quenching in dense environments;} considering that the majority of galaxies reside in groups or clusters (\citealt{Coil08} find that green valley galaxies are just as clustered as red sequence galaxies), it is not surprising that the majority of our galaxies are considered intermediate in morphology (see Table~\ref{subs}) {\newchange and therefore are undergoing or have undergone such an interaction.}

\subsection{Slow Quenching Timescales}\label{slow}
Although intermediate and rapid quenching timescales are the dominant mechanisms across the colour-magnitude diagram, together they cannot completely account for the quenching of disc galaxies. S14 concluded that slow quenching timescales were the most dominant mechanism for disc galaxies. {\changed However we show that: (i) intermediate quenching timescales are equally important in the green valley and (ii) rapid quenching timescales are equally important for {\tworef red galaxies with NUV emission}.} There is also a significantly lower preference for smooth galaxies to undergo such slow quenching timescales; suggesting that the evolution (or indeed creation) of typical smooth galaxies is dominated by processes external to the galaxy. {\changed This is excepting galaxies in the blue cloud where a small amount of slow evolution of blue ellipticals is occurring, presumably after a reinvigoration of star formation which is slowly depleting the gas available according to the Kennicutt$-$Schmidt law.}

\citet{Bamford09} using GZ1 vote fractions of galaxies in the SDSS, found a significant fraction of high stellar mass red spiral galaxies in the field. As these galaxies are isolated from the effects of interactions from other galaxies, the slow quenching mechanisms present in their preferred star formation histories are most likely due to secular processes (i.e. mechanisms internal to the galaxy, in the absence of sudden accretion or merger events; \citealt{KK04, Sheth12}). Bar formation in a disc galaxy is such a mechanism, whereby gas is funnelled to the centre of the galaxy by the bar over long timescales where it is used for star formation {\newchange \citep{Masters12, Saint12, Cheung13}}, consequently forming a `pseudo-bulge' \citep{Kormendy10, Simmons13}.

 If we believe that these slow quenching timescales are due to secular evolution processes, this is to be expected since these processes do not change the disc dominated nature of a galaxy. 

\subsection{Future Work}\label{future}
Due to the flexibility of our model we believe that the \starpy ~module will have a significant number of future applications, including the investigation of various different SFHs (e.g. constant SFR and starbursts). Considering the number of magnitude bands available across the SDSS, further analysis will also be possible with a larger set of optical and NUV colours, providing further constraints {\refchange and to ensure a more complete sample, containing a larger fraction of typical red sequence galaxies, if the need for NUV photometry was replaced with another band}. It would also be of interest to consider galaxies at higher redshift {\changed (e.g. out to $z \sim 1$ with Hubble Space Telescope photometry and the GZ:Hubble project, see \citealt{Melvin14} for first results) and consider different redshift bins in order to study the build up of the red sequence with cosmic time. }

With further use of the robust, detailed GZ2 classifications, we believe that \starpy~ will be able to distinguish any statistical difference in the star formation histories of barred vs. non-barred galaxies. This will require a simple swap of $\{p_s, p_d\}$ with $\{p_{bar}, p_{no bar}\}$ from the available GZ2 vote fractions. We believe that this will aid in the discussion of whether bars act to quench star formation (by funnelling gas into the galaxy centre) or promote star formation (by causing an increase in gas density as it travels through the disc) both sides of which have been fiercely argued \citep{Masters11, Masters12, Sheth05, Ellison11}.

{\newchange Further application of the \starpy ~code could be to investigate the SFH parameters of:
\begin{enumerate}
\item Currently merging/interacting pairs in comparison to those galaxies classified as merger remnants, from their degree of morphological disturbance,
\item Slow rotators and fast rotators which are thought to result from dry major mergers on the red sequence \citep{Em11} and gas rich, wet major mergers \citep{Em07} respectively,
\item Field and cluster galaxies using the projected neighbour density, $\Sigma$, from \citet{Baldry06}.
\end{enumerate}
}

\section{Conclusion}\label{conc}
We have used morphological classifications from the Galaxy Zoo 2 project to determine the morphology-dependent star formation histories of galaxies via a Bayesian analysis of an exponentially declining star formation quenching model. We determined the most likely parameters for the quenching onset time, $t_q$ and quenching timescale $\tau$ in this model for galaxies across the blue cloud, green valley and red sequence to trace galactic evolution across the colour-magnitude diagram. We find that the green valley is indeed a transitional population for all morphological types (in agreement with \citet{Sch2014}), however this transition proceeds slowly for the majority of disc-like galaxies and occurs rapidly for the majority of smooth-like galaxies in the red sequence. However, in addition to \citet{Sch2014}, {\changed our Bayesian approach has revealed a more nuanced result, specifically that the prevailing mechanism across all morphologies and populations is quenching with intermediate timescales}. Our main findings are as follows:
\begin{enumerate}
\item {\tworef The subset of red sequence galaxies with NUV emission studied in this investigation} are found to have similar preferences for quenching timescales compared to the green valley galaxies, but occurs at earlier quenching times regardless of morphology (see Figures~\ref{red_s} and~\ref{green_v}). Therefore the quenching mechanisms currently occurring in the green valley were also active in creating {\tworef the `blue end of of the red sequence'} at earlier times; confirming that the green valley is indeed a transitional population, regardless of morphology.

\item {\refchange We confirm that the typical {\tworef red galaxy with NUV emission studied in this investigation}, is elliptical in morphology and conclude that it has undergone a rapid to intermediate quench at some point in cosmic time, resulting in a very low current SFR (see Section~\ref{rs}.}

\item {\changed The green valley as it is currently observed is dominated by very slowly evolving disc-like galaxies along with intermediate- and smooth-like galaxies which pass across it with intermediate timescales within $\sim 1.0-1.5~\rm{Gyr}$ (see Section~\ref{gv}).}

\item {\refchange There are many different timescales responsible for quenching, causing a galaxy to progress through the green valley, which are dependant on galaxy type, with the smooth- and disc-like galaxies each having different dominant star formation histories across the colour-magnitude diagram. These timescales can be roughly split into three main regimes; rapid ($\tau < 1.0~$Gyr), intermediate ($1.0 < \tau~$[Gyr]~$< 2.0$) and slow ($\tau > 2.0~$ Gyr) quenching. }

\item {\changed Blue cloud galaxies are not well fit by a quenching model of star formation due to the continuous high star formation rates occurring (see Figure~\ref{blue_c}).}

\item {\changed Rapid quenching timescales are detected with a lower probability for green valley galaxies {\tworef than the subset of red sequence galaxies studied}.} We speculate that this quenching mechanism is caused by major mergers with black hole feedback, which are able to expel the remaining gas not initially exhausted in the merger-induced starburst and which can cause a change in morphology from disc- to bulge-dominated. The colour-change timescales from previous simulations of such events agree with our derived timescales  {\changed (see Section~\ref{rapid}). These rapid timescales are instrumental in forming red galaxies, however galaxies at the current epoch passing through the green valley do so at more intermediate timescales (see Figure~\ref{green_v}).}

\item Intermediate quenching timescales ($1.0 < ~\tau~\rm{[Gyr]}~ < 2.0 $) are found with constant probability across red and green galaxies for both smooth- and disc-like morphologies, the timescales for which agree with observed and simulated minor merger timescales (see Section~\ref{int}). We hypothesise such timescales can be caused by a number of external processes, including gas rich major mergers, mergers without black hole feedback, galaxy harassment, interactions and ram pressure stripping. The timescales and observed morphologies from previous studies agree with our findings, including that this is the dominant mechanisms for intermediate galaxies such as early-type spiral galaxies with spiral features but a dominant bulge, which split the GZ2 vote fractions (see Section~\ref{int}). 

\item Slow quenching timescales are the most dominant mechanism in the disc galaxy populations across the colour-magnitude diagram. Disc galaxies are often found in the field, therefore we hypothesise that such slow quenching timescales are caused by secular evolution and processes internal to the galaxy (see Section \ref{bc}). {\changed We also detect a small amount of slow quenching timescales for blue elliptical galaxies which we {\refchange attribute to a reinvigoration of star formation, the peak of which has passed and has started to decline by slowly depleting the gas available (see Section~\ref{bc}).}}

\item Due to the flexibility of this model we believe that the \starpy ~module compiled for this investigation will have a significant number of future applications, including the different star formation histories of barred vs non-barred galaxies, merging vs merger remnants, fast vs slow rotating elliptical galaxies and cluster vs field galaxies (see Section~\ref{future}).
\end{enumerate}

\section*{Acknowledgements}

The authors would like to thank the anonymous referee for helpful and insightful comments which improved both the presentation and the discussion of the results presented in this paper.

The authors would like to thank D. Forman-Mackey for extremely useful Bayesian statistics discussions, J. Binney for an interesting discussion on the nature of quenching and feedback in disc galaxies and M. Urry for the assistance in seeing the big picture. 

RS acknowledges funding from the Science and Technology Facilities Council Grant Code ST/K502236/1. BDS gratefully acknowledges support from the Oxford Martin School, Worcester College and Balliol College, Oxford. KS gratefully acknowledges support from Swiss National Science Foundation Grant PP00P2\_138979/1. KLM acknowledges funding from The Leverhulme Trust as a 2010 Early Career Fellow. TM acknowledges funding from the Science and Technology Facilities Council Grant Code ST/J500665/1. KWW and LF acknowledge funding from a Grant-in-Aid from the University of Minnesota.

The development of Galaxy Zoo was supported in part by the Alfred P. Sloan Foundation. Galaxy Zoo was supported by The Leverhulme Trust. 

Based on observations made with the NASA Galaxy Evolution Explorer.  GALEX is operated for NASA by the California Institute of Technology under NASA contract NAS5-98034

Funding for the SDSS and SDSS-II has been provided by the Alfred P. Sloan Foundation, the Participating Institutions, the National Science Foundation, the U.S. Department of Energy, the National Aeronautics and Space Administration, the Japanese Monbukagakusho, the Max Planck Society, and the Higher Education Funding Council for England. The SDSS Web Site is \url{http://www.sdss.org/}.
The SDSS is managed by the Astrophysical Research Consortium for the Participating Institutions. The Participating Institutions are the American Museum of Natural History, Astrophysical Institute Potsdam, University of Basel, University of Cambridge, Case Western Reserve University, University of Chicago, Drexel University, Fermilab, the Institute for Advanced Study, the Japan Participation Group, Johns Hopkins University, the Joint Institute for Nuclear Astrophysics, the Kavli Institute for Particle Astrophysics and Cosmology, the Korean Scientist Group, the Chinese Academy of Sciences (LAMOST), Los Alamos National Laboratory, the Max-Planck-Institute for Astronomy (MPIA), the Max-Planck-Institute for Astrophysics (MPA), New Mexico State University, Ohio State University, University of Pittsburgh, University of Portsmouth, Princeton University, the United States Naval Observatory, and the University of Washington.

This publication made extensive use of the Tool for Operations on Catalogues And Tables (TOPCAT; ~\citealt{Taylor05}) which can be found at \url{http://www.star.bris.ac.uk/~mbt/topcat/}. Ages were calculated from the observed redshifts using the \emph{cosmolopy} package provided in the Python module \emph{astroPy}\footnote{\url{http://www.astropy.org/}}; \citealt{Rob13}). This research has also made use of NASA's ADS service and Cornell's ArXiv. 

{}

\appendix

\section{Testing starpy}\label{app_test}

\begin{figure*}
\centering{
\includegraphics[width=\textwidth]{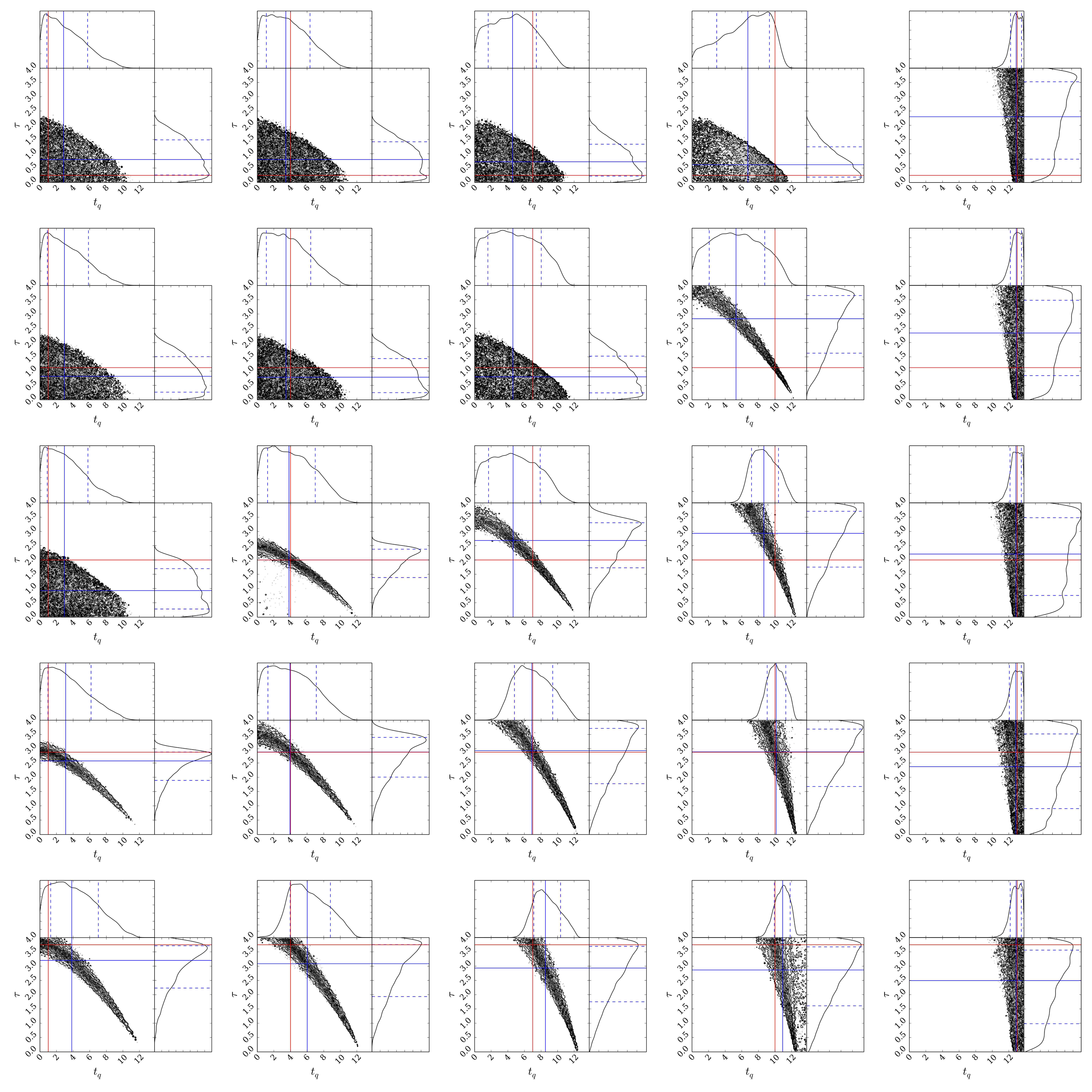}}
\caption{Results from \starpy ~for an array of synthesised galaxies with known, i.e. \underline{true}, $t_q$ and $\tau$ values (marked by the red lines) using the complete function to calculate the predicted colour of a proposed set of $\theta$ values in each MCMC iteration, assuming an error on the calculated known colours of $\sigma_{u-r} = 0.124$ and $\sigma_{NUV-u} = 0.215$, the average errors on the GZ sample colours. In each case \starpy ~succeeds in locating the true parameter values within the degeneracies of the star formation history model. These degeneracies can clearly be seen in Figure~\ref{pred}.}
\label{test_mosaic}
\end{figure*}

In order to test that \starpy ~can find the correct quenching model for a given observed colour, {\refchange 25 synthesised galaxies were created with known SFHs (i.e. known values of $\theta$) from which optical and NUV colours were generated. These were input into \starpy ~ to ensure that the known values of $\theta$ were reproduced, within error, for each of the 25 synthesised galaxies. Figure~\ref{test_mosaic} shows the results for each of these 25 synthesised galaxies}, with the known values of $\theta$ shown by the red lines. In some cases this red line does not coincide with the peak of the distribution shown in the histograms for one parameter, however in all cases the intersection of the red lines is within the sample contours. 

{\refchange We find peaks in the histograms across all areas of the parameter space in both dimensions of $[t, \tau]$, this ensures that the results presented in Figures~\ref{red_s},~\ref{green_v} \&~\ref{blue_c} arise due to a superposition of extended probability distributions, as opposed to a bimodal distribution of probability distributions across all galaxies.}

\section{Using look up tables}\label{app_lookup}

\begin{figure*}
\centering{
\includegraphics[width=0.48\textwidth]{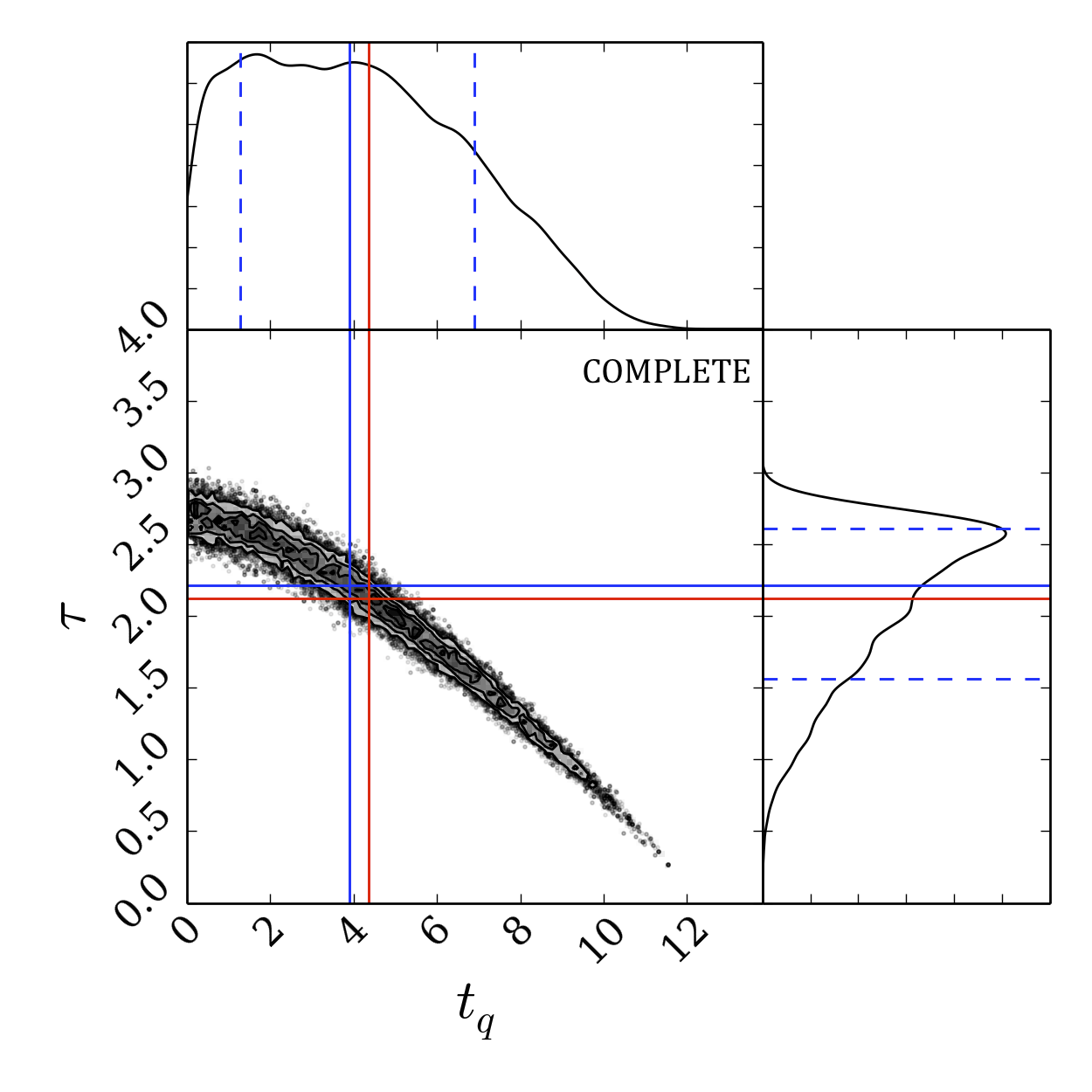}
\includegraphics[width=0.48\textwidth]{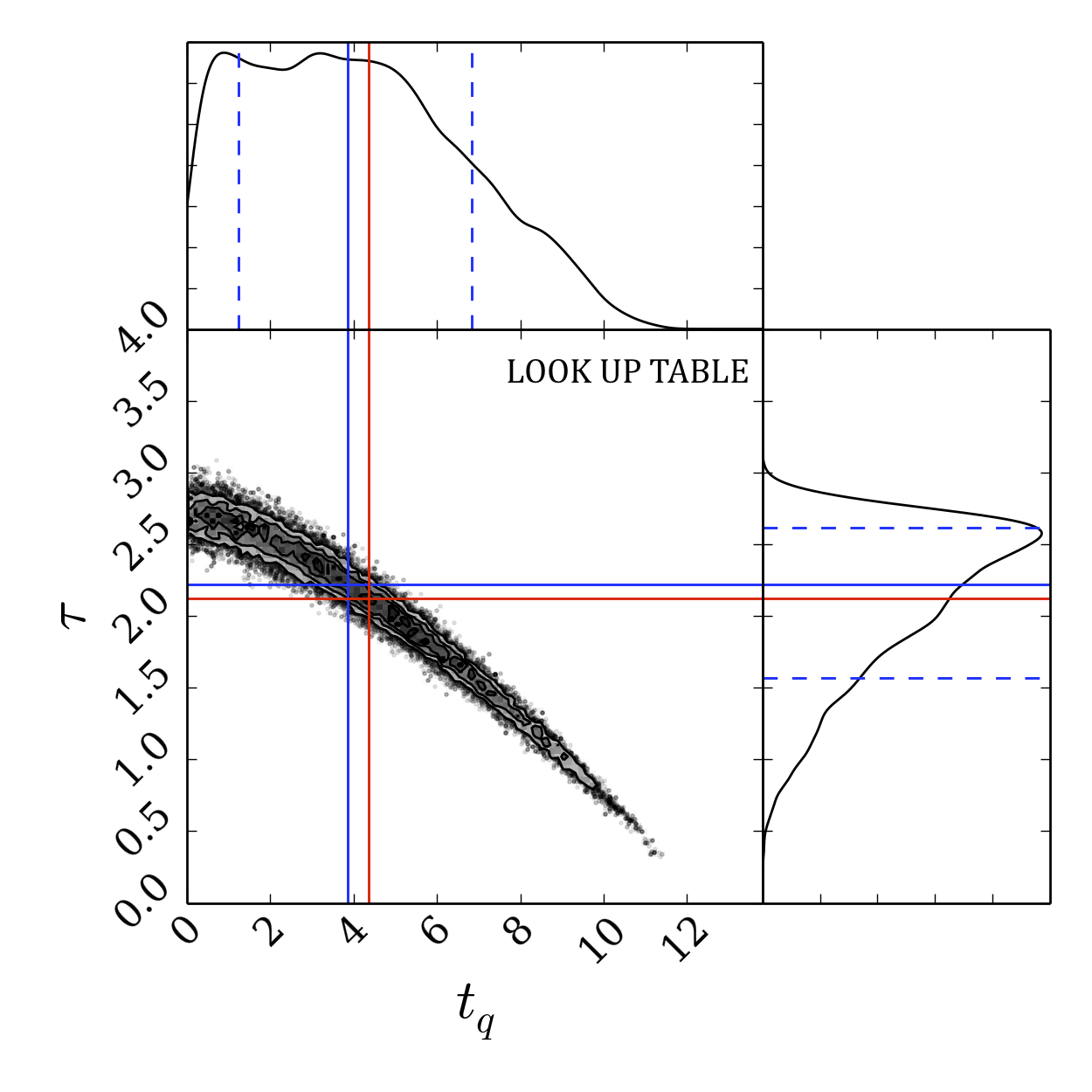}}
\caption{Left panel: Results from \starpy ~for \underline{true} $t_q$ and $\tau$ values (red lines) using the complete function to calculate the predicted colour of a proposed set of $\theta$ values in each MCMC iteration. The median walker position {\refchange(the 50th percentile of the Bayesian probability distribution)} is shown by the solid blue line with the dashed lines encompassing $68\% (\pm 1\sigma)$ of the samples {\refchange(the 16th and 84th percentile positions)}. The time taken to run for a single galaxy using this method is approximately 2 hours. Right panel: Results from \starpy ~for \underline{true} $t_q$ and $\tau$ values using a look up table generated from the complete function to calculate the predicted colour of a proposed set of $\theta$ values in each MCMC iteration. The time taken to run for a single galaxy using this method is approximately 2 minutes.}
\label{lookup}
\end{figure*}

\begin{table*}
\centering{
\caption{{\refchange Median walker positions {\refchange(the 50th percentile; as shown by the blue solid lines in Figure~\ref{lookup})} found by \starpy ~ for a single galaxy}, using the complete star formation history function and a look up table to speed up the run time. The errors quoted define the region in which $68\%$ of the samples are located, shown by the dashed blue lines in Figure~\ref{lookup}. The known true values are also quoted, as shown by the red lines in Figure~\ref{lookup}. All values are quoted to three significant figures.}
\begin{tabular*}{0.65\textwidth}{r @{\extracolsep{\fill}}ccc}
\multicolumn{1}{l}{} & \multicolumn{3}{c}{}                                          \\ \hline
                     & $t_q$                       & $\tau$                       &  \\ \hline
True                 & $4.37$                        & $2.12$                         &  \\
Complete             & $3.893 \pm^{3.014}_{2.622}$ & $2.215 \pm^{0.395}_{0.652}$ &  \\
Look up table        & $3.850 \pm^{2.988}_{2.619}$ & $2.218 \pm^{0.399}_{0.649}$ & \\ \hline
\end{tabular*}}
\label{median_lu}
\end{table*}

{\newchange Considering the size of the sample in this investigation of $126,316$ galaxies total, a three dimensional look up table (in observed time, quenching time and quenching rate) was generated using the star formation history function in \starpy ~to speed up the run time. Figure~\ref{lookup} shows an example of how using the look up table in place of the full function does not affect the results to a significant level. Table~\ref{median_lu} quotes the median walker positions {\refchange (the 50th percentile of the Bayesian probability distribution) }along with their $\pm 1\sigma$ ranges for both methods in comparison to the true values specified to test \starpy. The uncertainties incorporated into the quoted values by using the look up table are therefore minimal with a maximum $\Delta = 0.043$.}

{\refchange \section{Discarding Poorly Fit Galaxies}\label{discard}

\begin{table*}
\centering{
\caption{{\refchange Number of galaxies in each population which had walker positions discarded due to low probability in order to exclude those galaxies from the analysis which were poorly fit by this quenching model.}}
\begin{tabular*}{0.85\textwidth}{r @{\extracolsep{\fill}} ccc}
                                          & \textbf{Red Sequence}                                   & \textbf{Green Valley}                                  & \textbf{Blue Cloud}                                      \\ \hline
All walkers discarded                     & \begin{tabular}[c]{@{}c@{}}1420\\ (7.00\%)\end{tabular} & \begin{tabular}[c]{@{}c@{}}437\\ (2.41\%)\end{tabular} & \begin{tabular}[c]{@{}c@{}}3109\\ (5.37\%)\end{tabular}  \\
More than half walker positions discarded & \begin{tabular}[c]{@{}c@{}}2010\\ (9.92\%)\end{tabular} & \begin{tabular}[c]{@{}c@{}}779\\ (4.30\%)\end{tabular} & \begin{tabular}[c]{@{}c@{}}6669\\ (11.52\%)\end{tabular} \\ \hline
\end{tabular*}}
\label{discardnum}
\end{table*}

We discard walker positions returned by \starpy~ with a corresponding probability of $P(\theta_k|d_k) < 0.2$ in order to exclude galaxies which are not well fit by the quenching model; for example blue cloud galaxies which are still star forming will be poorly fit by a quenching model (see Section~\ref{qmod}). This raises the issue of whether we exclude a significant fraction of our galaxy sample and whether those galaxies reside in a specific location of the colour-magnitude. The fraction of galaxies which had all or more than half of their walker positions discarded due to low probability are shown in Table \ref{discardnum}.

This is not a significant fraction of either population, therefore this shows that the \starpy~ module is effective in fitting the majority of galaxies and that this method of discarding walker positions ensures that poorly fit galaxies are removed from the analysis of the results. Figure \ref{discarded} shows that these galaxies with discarded walker positions are also scattered across the optical-NUV colour-colour diagram and therefore \starpy ~is also effective in fitting galaxies across this entire plane. 

\begin{figure*}
\includegraphics[width=0.9\textwidth]{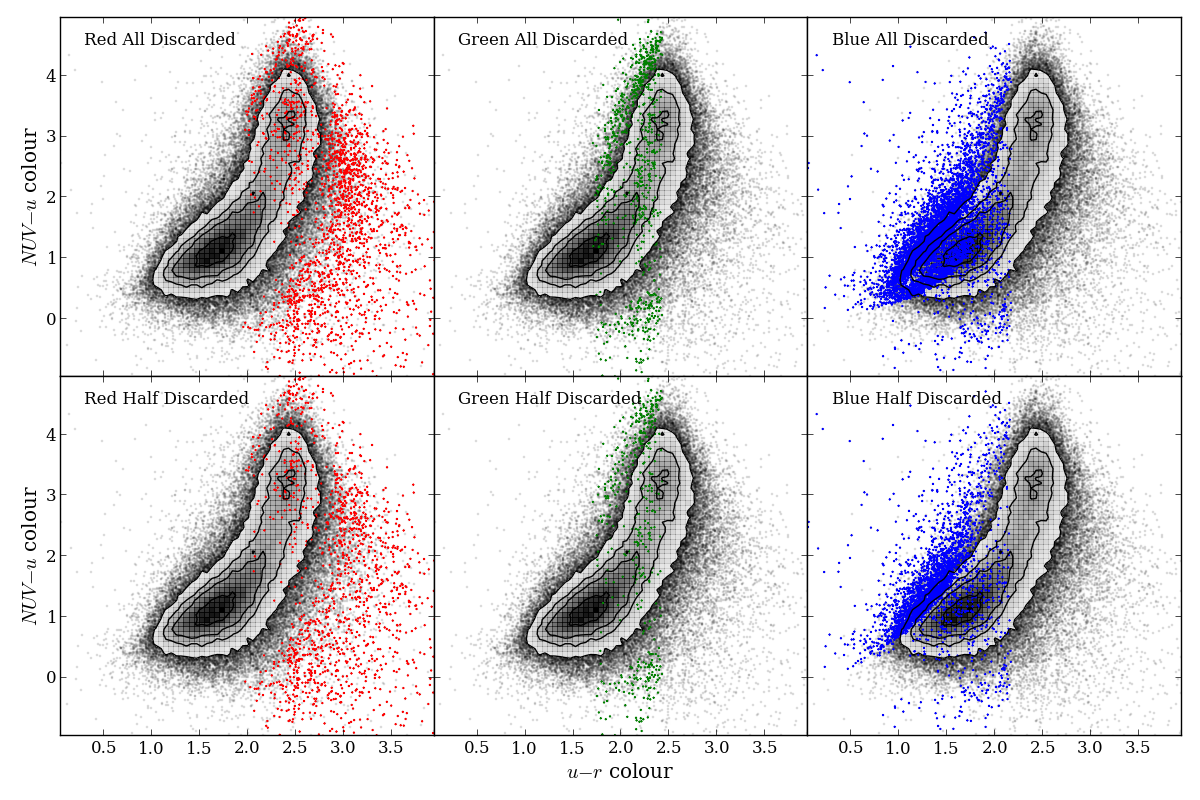}
\caption{Contours show the full GZ2 subsample optical-NUV colour-colour diagram. The points show the positions of the galaxies which had all (top panels) or more than half (bottom panel) of their walker positions discarded due to their low probability for the red sequence (left), green valley (middle) and blue cloud (right).}
\label{discarded}
\end{figure*}
}

{\refchange \section{Observability of quenching galaxies} \label{observe}

The numbers of galaxies found undergoing a rapid quench will be underestimated compared to the true value due to their observability, i.e. their time spent in the green valley is extremely short, so detecting a galaxy there is difficult. We considered this time spent in the green valley across our model parameter space of star formation histories and the results are shown below in Figure \ref{obsplot}.
\begin{figure*}
\centering{
\includegraphics[width=0.45\textwidth]{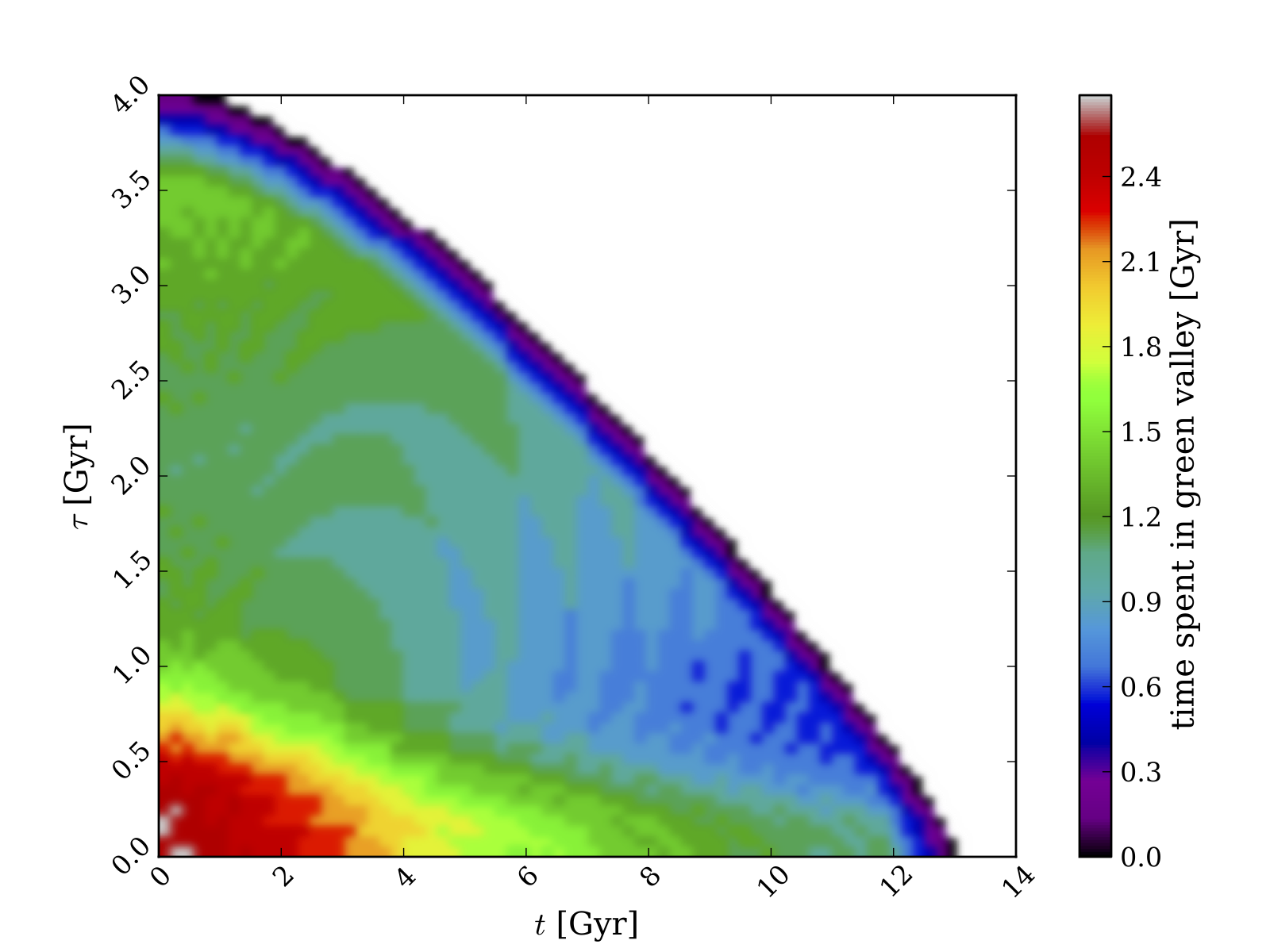}}
\caption{Plot showing the time spent in the green valley across the model star formation history parameter space. This affects the observability of those galaxies which have quenched rapidly and recently and have passed too quickly through the green valley to be detected. The white region denotes those models with colours that do not enter the green valley by the present cosmic time.}
\label{obsplot}
\end{figure*}
}

\end{document}